\documentstyle[11pt]{article}
\title{Evolutionary game theory and population dynamics} 
\author{Jacek Mi\c{e}kisz \\ Institute of Applied Mathematics \\
and Mechanics \\ Warsaw University  \\ ul. Banacha 2  \\ 02-097
Warsaw, Poland \\ e-mail: miekisz@mimuw.edu.pl} 
\begin{document} 
\baselineskip=18pt
\maketitle 

\maketitle

\newtheorem{theo}{Theorem}
\newtheorem{defi}{Definition}
\newtheorem{hypo}{Hypothesis}

\section{Short overview}

We begin these lecture notes by a crash course in game theory. 
In particular, we introduce a fundamental notion of a Nash equilibrium. 
To address the problem of the equilibrium selection in games with multiple equilibria, 
we review basic properties of the deterministic replicator dynamics 
and stochastic dynamics of finite populations. 

We show the almost global asymptotic stability of an efficient equilibrium 
in the replicator dynamics with a migration between subpopulations. 
We also show that the stability of a mixed equilibrium depends on the time delay 
introduced in replicator equations. For large time delays, a population oscillates 
around its equilibrium.

We analyze the long-run behaviour of stochastic dynamics in well-mixed populations 
and in spatial games with local interactions. We review results concerning the effect 
of the number of players and the noise level on the stochastic stability of Nash equilibria. 
In particular, we present examples of games in which when the number of players increases 
or the noise level decreases, a population undergoes a transition between its equilibria. 
We discuss similarities and differences between systems of interacting players in spatial games 
maximizing their individual payoffs and particles in lattice-gas models minimizing 
their interaction energy. 

In short, there are two main themes of our lecture notes: the selection of efficient equilibria
(providing the highest payoffs to all players) in population dynamics and the dependence 
of the long-run behaviour of a population on various parameters such as the time delay, 
the noise level, and the size of the population.

\section{Introduction}

\noindent Many socio-economic and biological processes can be modeled as systems 
of interacting individuals; see for example econophysics bulletin \cite{ekono} 
and statistical mechanics and quantitative biology archives \cite{quantbiol}. 
One may then try to derive their global behaviour from individual interactions 
between their basic entities such as animals in ecological and evolutionary models, 
genes in population genetics and people in social processes. Such approach is fundamental 
in statistical physics which deals with systems of interacting particles. 
One can therefore try to apply methods of statistical physics to investigate 
the population dynamics of interacting individuals. There are however profound differences 
between these two systems. Physical systems tend in time to states which are characterized 
by the minimum of some global quantity, the total energy or free energy of the system. 
Population dynamics lacks such general principle. Agents in social models maximize 
their own payoffs, animals and genes maximize their individual darwinian fitness. 
The long-run behavior of such populations cannot in general be characterized by the global 
or even local maximum of the payoff or fitness of the whole population. 
We will explore similarities and differences between these systems.
     
The behaviour of systems of interacting individuals can be often described within 
game-theoretic models \cite{maynard3,fuden1,fuden2,wei,vr,samuel,hof2,young2,gintis,cress,ams,nowak,nowakscience,nowakbook}. 
In such models, players have at their disposal certain strategies 
and their payoffs in a game depend on strategies chosen both by them 
and by their opponents. The central concept in game theory 
is that of a {\bf Nash equilibrium}. It is an assignment of strategies to players 
such that no player, for fixed strategies of his opponents, has an incentive to deviate 
from his current strategy; no change can increase his payoff. 

In Chapter 3, we present a crash course in game theory.
One of the fundamental problems in game theory is the equilibrium selection
in games with multiple Nash equilibria. Some two-player symmetric games with two strategies, 
have two Nash equilibria and it may happen that one of them is payoff dominant 
(also called efficient) and the other one is risk-dominant. In the efficient equilibrium, players 
receive highest possible payoffs. The strategy is risk-dominant 
if it has a higher expected payoff against a player playing both strategies 
with equal probabilities. It is played by individuals averse to risk. 
One of the selection methods is to construct a dynamical system 
where in the long run only one equilibrium is played with a high frequency. 

John Maynard Smith \cite{maynard1,maynard2,maynard3} has refined the concept of the Nash equilibrium to include 
the stability of equilibria against mutants. He introduced the fundamental notion 
of an {\bf evolutionarily stable strategy}. If everybody plays such a strategy, then the small number 
of mutants playing a different strategy is eliminated from the population. 
The dynamical interpretation of the evolutionarily stable strategy was later provided 
by several authors \cite{tayjon,hof1,zee}. They proposed a system of difference 
or differential replicator equations which describe the time-evolution of frequencies of strategies. 
Nash equilibria are stationary points of this dynamics. It appears that in games with 
a payoff dominant equilibrium and a risk-dominant one, both are asymptotically stable 
but the second one has a larger basin of attraction in the replicator dynamics.

In Chapter 4, we introduce {\bf replicator dynamics} and review theorems concerning 
asymptotic stability of Nash equilibria \cite{wei,hof2,ams}. 
Then in Chapter 5, we present our own model of the replicator dynamics \cite{migration} 
with a migration between two subpopulations for which an efficient equilibrium is  
almost globally asymptotically stable. 

It is very natural, and in fact important, to introduce a {\bf time delay} 
in the population dynamics; a time delay between acquiring information 
and acting upon this knowledge or a time delay between playing games and receiving payoffs.
Recently Tao and Wang \cite{taowang} investigated the effect of a time delay 
on the stability of interior stationary points of the replicator dynamics. 
They considered two-player games with two strategies and a unique asymptotically 
stable interior stationary point. They proposed a certain form of a time-delay 
differential replicator equation. They showed that the mixed equilibrium 
is asymtotically stable if a time delay is small. 
For sufficiently large delays it becomes unstable. 

In Chapter 6, we construct two models of discrete-time replicator dynamics
with a time delay \cite{delay}. In the social-type model, players imitate opponents 
taking into account average payoffs of games played some units of time ago.
In the biological-type model, new players are born from parents who played
in the past. We consider two-player games with two strategies and a unique mixed 
Nash equilibrium. We show that in the first type of dynamics,
it is asymptotically stable for small time delays and becomes unstable for large ones
when the population oscillates around its stationary state. 
In the second type of dynamics, however, the Nash equilibrium 
is asymptotically stable for any time delay. Our proofs are elementary, 
they do not rely on the general theory of delay differential and difference equations.

Replicator dynamics models population behaviour in the limit of the infinite number of individuals.
However, real populations are finite. Stochastic effects connected with random matchings of players, 
mistakes of players and biological mutations can play a significant role in such systems. 
We will discuss various stochastic adaptation dynamics of populations with a fixed number of players 
interacting in discrete moments of time. In well-mixed populations, individuals are randomly matched 
to play a game \cite{kmr,rvr,population}. The deterministic selection part of the dynamics ensures 
that if the mean payoff of a given strategy is bigger than the mean payoff of the other one, 
then the number of individuals playing the given strategy increases. 
However, players may mutate hence the population may move against 
a selection pressure. In spatial games, individuals are located 
on vertices of certain graphs and they interact only with their neighbours; see for example 
\cite{nowak0,nowak1,nowak2,blume1,ellis1,young2,ellis2,linnor,doebeli1,doebeli2,szabo1,szabo2, szabo6,hauert1,doebeli3,hauert2,hauert3} and a recent review \cite{szabo8} and references therein. 
In discrete moments of times, players adapt to their opponents 
by choosing with a high probability the strategy which is the best response, 
i.e. the one which maximizes the sum of the payoffs obtained from individual games. 
With a small probability, representing the noise of the system, they make mistakes. 
The above described stochastic dynamics constitute ergodic Markov chains with states 
describing the number of individuals playing respective strategies or corresponding to complete profiles
of strategies in the case of spatial games. Because of the presence of random mutations, 
our Markov chains are ergodic (irreducible and periodic) and therefore they possess unique stationary measures. 
To describe the long-run behavior of such stochastic dynamics, Foster and Young \cite{foya} introduced 
a concept of stochastic stability. A configuration of the system is {\bf stochastically stable} 
if it has a positive probability in the stationary measure of the corresponding Markov chain 
in the zero-noise limit, that is the zero probability of mistakes. It means that in the long run 
we observe it with a positive frequency along almost any time trajectory. 

In Chapter 7, we introduce the concept of stochastic stability and present 
a useful representation of stationary measures of ergodic Markov chains
\cite{freiwen1,freiwen2,shub}.
 
In Chapter 8, we discuss populations with random matching of players 
in {\bf well-mixed populations}. We review recent results concerning 
the dependence of the long-run behavior of such systems on the number of players 
and the noise level. In the case of two-player games with two symmetric Nash equilibria, 
an efficient one and a risk-dominant one, when the number of players increases, 
the population undergoes twice a transition between its equilibria.  
In addition, for a sufficiently large number of individuals, 
the population undergoes another {\bf equilibrium transition} 
when the noise decreases.   

In Chapter 9, we discuss {\bf spatial games}. We will see that in such models, 
the notion of a Nash equilibrium (called there a Nash configuration) is similar to the notion 
of a ground-state configuration in classical lattice-gas models of interacting particles. 
We discuss similarities and differences between systems of interacting players in spatial games 
maximizing their individual payoffs and particles in lattice-gas models minimizing 
their interaction energy. 

The concept of stochastic stability is based on the zero-noise limit 
for a fixed number of players. However, for any arbitrarily low but fixed noise, 
if the number of players is large enough, the probability of any individual configuration 
is practically zero. It means that for a large number of players, 
to observe a stochastically stable configurations we must assume that players 
make mistakes with extremely small probabilities. On the other hand, it may happen that 
in the long run, for a low but fixed noise and sufficiently large number 
of players, the stationary configuration is highly concentrated on an ensemble
consisting of one Nash configuration and its small perturbations, 
i.e. configurations where most players play the same strategy.
We will call such configurations {\bf ensemble stable.}
It will be shown that these two stability concepts do not necessarily coincide.
We will present examples of spatial games with three strategies 
where concepts of stochastic stability and ensemble stability do not coincide 
\cite{statmech,physica}. In particular, we may have the situation, 
where a stochastically stable strategy is played in the long run with an arbitrarily low frequency. 
In fact, when the noise level decreases, the population undergoes a sharp transition 
with the coexistence of two equilibria for some noise level. 
Finally, we discuss the influence of {\bf dominated strategies} on the long-run behaviour 
of population dynamics.

In Chapter 10, we shortly review other results concerning stochastic dynamics of finite populations.
 
\section{A crash course in game theory}

To characterize a game-theoretic model one has to specify players, 
strategies they have at their disposal and payoffs they receive.  
Let us denote by $I=\{1,...,n\}$ the set of players.
Every player has at his disposal $m$ different strategies. Let $S=\{1,...,m\}$
be the set of strategies, then $\Omega=S^{I}$ is the set of strategy profiles, 
that is functions assigning strategies to players. The payoff of any player 
depends not only on his strategy but also on strategies of all other players.  
If $X \in \Omega$, then we write $X=(X_{i}, X_{-i})$, where $X_{i} \in S$ 
is a strategy of the i-th player and $X_{-i} \in S^{I-\{i\}}$ is a strategy 
profile of remaining players. The payoff of the i-th player is a function defined 
on the set of profiles,

$$U_{i}: \Omega \rightarrow R, \; \; i,...,n$$  

The central concept in game theory is that of a Nash equilibrium. 
An assignment of strategies to players is a {\bf Nash equilibrium}, 
if for each player, for fixed strategies of his opponents, 
changing his current strategy cannot increase his payoff. 
The formal definition will be given later on when we enlarge 
the set of strategies by mixed ones.

Although in many models the number of players is very large 
(or even infinite as we will see later on in replicator dynamics models), 
their strategic interactions are usually decomposed into a sum of two-player games. 
Only recently, there have appeared some systematic studies of truly multi-player games 
\cite{kim,broom,multi,tplatk3}. Here we will discuss only two-player games with two or three strategies.
We begin with games with two strategies, $A$ and $B$. Payoffs functions can be then represented 
by $2 \times 2$ payoff matrices. A general payoff matrix is given by
\vspace{3mm}

\hspace{25mm} A  \hspace{4mm} B   
\vspace{1mm}

\hspace{15mm} A \hspace{5mm} a  \hspace{5mm} b 

U = \hspace{6mm} 

\hspace{15mm} B \hspace{5mm} c  \hspace{5mm} d,
\vspace{3mm}

where $U_{kl}$, $k,l = A, B$, is a payoff of the first (row) player when
he plays the strategy $k$ and the second (column) player plays the strategy $l$.
\vspace{1mm}
 
We assume that both players are the same and hence payoffs of the column player are given 
by the matrix transposed to $U$; such games are called symmetric. In this classic set-up
of static games (called matrix games or games in the normal form), 
players know payoff matrices, simultaneously announce (use) their strategies 
and receive payoffs according to their payoff matrices.  

We will present now three main examples of symmetric two-player games with two strategies.
We begin with an anecdote, then an appropriate game-theoretic model is build 
and its Nash equilibria are found.
\vspace{4mm}

\noindent {\bf Example 1 (Stag-hunt game)}
\vspace{3mm}

\noindent Jean-Jacques Rousseau wrote, in his Discourse on the Origin and Basis of Equality among Men,
about two hunters going either after a stag or a hare \cite{ordershook,fuden1}. In order to get a stag, 
both hunters must be loyal one to another and stay at their positions. 
A single hunter, deserting his companion, can get his own hare. In the game-theory language,
we have two players and each of them has at his disposal two strategies: Stag (St) and Hare (H).
In order to present this example as a matrix game we have to assign some values to animals. 
Let a stag (which is shared by two hunters) be worth 10 units and a hare 3 units. 
Then the payoff matrix of this symmetric game is as follows:
\vspace{3mm}

\hspace{23mm} St  \hspace{4mm} H   
\vspace{1mm}

\hspace{15mm} St \hspace{3mm} 5  \hspace{5mm} 0 

U = \hspace{6mm} 

\hspace{15mm} H \hspace{4mm} 3  \hspace{5mm} 3
\vspace{4mm}

\noindent It is easy to see that there are two Nash equilibria: $(St,St)$ and $(H,H)$.

In a general payoff matrix, if $a>c$ and $d>b$, then both $(A,A)$ and $(B,B)$ are Nash equilibria. 
If $a+b<c+d$, then the strategy $B$ has a higher expected payoff against a player playing 
both strategies with the probability $1/2$. We say that $B$ risk dominates the strategy $A$ 
(the notion of the risk-dominance was introduced and thoroughly studied 
by Hars\'{a}nyi and Selten \cite{hs}). If at the same time $a>d$, 
then we have a selection problem of choosing between the payoff-dominant 
(Pareto-efficient) equilibrium $(A,A)$ and the risk-dominant $(B,B)$.
\vspace{4mm}

\noindent {\bf Example 2 (Hawk-Dove game)}
\vspace{3mm}

\noindent Two animals are fighting for a certain territory of a value V. 
They can be either aggressive (hawk strategy - H) or peaceful (dove strategy - D).
When two hawks meet, they accure the cost of fighting $C>V$ and then they split the territory.
When two dove meets, they split the territory without a fight. 
A dove gives up the territory to a hawk. We obtain the following payoff matrix:
\eject

\hspace{28mm} H  \hspace{11mm} D   
\vspace{1mm}

\hspace{15mm} H \hspace{3mm} (V-C)/2  \hspace{6mm} V 

U = \hspace{6mm} 

\hspace{15mm} D \hspace{7mm} 0  \hspace{11mm} V/2,
\vspace{4mm}

The Hawk-Dove game was analyzed by John Maynard Smith \cite{maynard3}. 
It is also known as the Chicken game \cite{russell} or the Snowdrift game \cite{doebeli3}.
It has two non-symmetric Nash equilibria: $(H,D)$ and $(D,H)$.
\vspace{3mm}

\noindent {\bf Example 3 (Prisoner's Dilemma)}
\vspace{3mm}

\noindent The following story was discussed by Melvin Dresher, Merill Flood, and Albert Tucker
\cite{axelrod,poundstone,sigmund}. Two suspects of a bank robbery are caught and interrogated by the police.
The police offers them separately the following deal. If a suspect testifies against  
his colleague (a strategy of defection - D), and the other does not (cooperation - C), 
his sentence will be reduced by five years. If both suspects testify, that is defect, 
they will get the reduction of only one year. However, if they both cooperate and do not testify, 
their sentence, because of the lack of a hard evidence, will be reduced by three years. 
We obtain the following payoff matrix:
\vspace{3mm}

\hspace{22mm} C  \hspace{5mm} D   
\vspace{1mm}

\hspace{15mm} C \hspace{3mm}  3  \hspace{5mm} 0 

U = \hspace{6mm} 

\hspace{15mm} D \hspace{3mm}  5  \hspace{5mm} 1
\vspace{4mm}

The strategy $C$ is a {\bf dominated strategy} - it results in a lower payoff 
than the strategy $D$, regardless of a strategy used by the other player.
Therefore, $(D,D)$ is the unique Nash equilibrium but both players 
are much better off when they play $C$ - this is the classic Prisoner's Dilemma. 
\vspace{3mm}

A novel behaviour can appear in games with three strategies.
\vspace{3mm}

\noindent {\bf Example 4 (Rock-Scissors-Paper game)}
\vspace{2mm}

\noindent In this game, each of two players simultaneously exhibits a sign of either a scissors ($S$), 
a rock ($R$), or a paper ($P$). The game has a cyclic behaviour: rock crashes scissors, 
scissors cut paper, and finally paper wraps rock. The payoffs can be given by the following matrix:
\vspace{5mm}

\hspace{22mm} R  \hspace{2mm} S \hspace{3mm} P  

\hspace{14mm} R  \hspace{3mm} 1  \hspace{3mm} 2 \hspace{3mm} 0

U = \hspace{6mm} S \hspace{4mm} 0  \hspace{3mm} 1 \hspace{3mm} 2

\hspace{14mm} P \hspace{3mm} 2  \hspace{3mm} 0 \hspace{3mm} 1
\vspace{3mm}

It is easy to verify that this game, because of its cyclic behavior, 
does not have any Nash equilibria as defined so far. However, we intuitively feel that
when we repeat it many times, the only way not to be exploited 
is to mix randomly strategies, i.e. to choose each strategy 
with the probability $1/3$. 
     
This brings us to a concept of a mixed stategy, a probability
mass function on the set of pure strategies $S$. 
Formally, a {\bf mixed strategy} $x$ is an element of a simplex
$$\Delta=\{x \in R^{m}, 0 \leq x_{k} \leq 1, \sum_{k=1}^{m} x_{k} =1\}.$$
By the support of a mixed strategy $x$ we mean the set of pure strategies 
with positive probabilities in $x$.
Payoffs of mixed strategies are defined as appropriate expected values.
In two-player games, a player who uses a mixed strategy $x$ against a player 
with a mixed strategy $y$ receives a payoff given by 
$$\sum_{k,l \in S}U_{kl}x_{k}y_{l}.$$
In general n-player games, profiles of strategies are now elements of $\Theta=\Delta^{I}.$   
We are now ready to define formally a Nash equilibrium.

\begin{defi}
$X \in \Theta$ is a {\bf Nash equilibrium} if for every $i \in I$
and every $y \in \Delta$, 
$$U_{i}(X_{i},X_{-i}) \geq U_{i}(y,X_{-i})$$ 
\end{defi}

In the mixed Nash equilibrium, expected payoffs of all strategies
in its support should be equal. Otherwise a player could increase his payoff by increasing 
the probability of playing a strategy with the higher expected payoff.
In two-player games with two strategies, we identify a mixed strategy 
with its first component, $x = x_{1}.$ Then the expected payoff of $A$ is given 
by $ax+b(1-x)$ and that of $B$ by $cx+d(1-x)$. $x^{*}= (d-b)/(d-b+a-c)$ for which 
the above two expected values are equal is a mixed Nash equilibrium or more formally,
a profile $(x,x)$ is a Nash equilibrium.

In Examples 1 and 2, in addition to Nash equilibria in pure strategies, we have
mixed equilibria, $x^{*}=3/5$ and $x^{*}=V/C$ respectively.  
It is obvious that the Prisoner's Dilemma game does not have any mixed Nash equilibria
.
On the other hand, the only Nash equilibrium of the Rock-Scissors-Paper game 
is a mixed one assigning the probability $1/3$ to each strategy.

We end this chapter by a fundamental theorem due to John Nash \cite{nash1,nash2}.

\begin{theo}
Every game with a finite number of players and a finite number of strategies
has at least one Nash equilibrium. 
\end{theo}

In any Nash equilibrium, every player uses a strategy which is a best reply 
to the profile of strategies of remaining players. Therefore a Nash equilibrium 
can be seen as a best reply to itself - a fixed point of a certain best-reply correspondence.
Then one can use the Kakutani fixed point theorem to prove the above theorem. 
\eject

\section{Replicator dynamics}

The concept of a Nash equilibrium is a static one. Here we will introduce 
the classical replicator dynamics and review its main properties \cite{wei,hof2,ams}. 
Replicator dynamics provides a dynamical way of achieving Nash equilibria in populations. 
We will see that Nash equilibria are stationary points of such dynamics and some of them 
are asymptotically stable. 

Imagine a finite but a very large population of individuals. Assume that they are paired
randomly to play a symmetric two-player game with two strategies and the payoff matrix given 
in the beginning of the previous chapter. The complete information about such population 
is provided by its strategy profile, that is an assignment of pure strategies to players.
Here we will be interested only in the proportion of individuals playing respective strategies. 
We assume that individuals receive average payoffs with respect to all possible opponents - 
they play against the average strategy.

Let $r_{i}(t)$, $i=A, B,$ be the number of individuals playing the strategy 
$A$ and $B$ respectively at the time $t$. Then  $r(t)=r_{A}(t)+r_{B}(t)$ is the total number of players
and $x(t)=\frac{r_{1}(t)}{r(t)}$ is a fraction of the population playing $A$.

We assume that during the small time interval $\epsilon$, only an $\epsilon$ fraction 
of the population takes part in pairwise competitions, that is plays games.
We write
\begin{equation}
r_{i}(t + \epsilon) = (1-\epsilon)r_{i}(t) + \epsilon r_{i}(t)U_{i}(t); \; \; i= A,B,
\end{equation}

where $U_{A}(t)= ax(t)+b(1-x(t))$ and $U_{B}(t)= cx(t)+d(1-x(t))$  
are average payoffs of individuals playing A and B respectively.
We assume that all payoffs are not smaller than $0$ hence $r_{A}$ and $r_{B}$
are always non-negative and therefore $0\leq x \leq 1$. 

The equation for the total number of players reads

\begin{equation}
r(t + \epsilon) = (1-\epsilon)r(t) + \epsilon r(t)\bar{U}(t),
\end{equation}
where $\bar{U}(t)=x(t)U_{A}(t)+(1-x(t))U_{B}(t)$ is the average payoff in the population at the time $t$.
When we divide (1) by (2) we obtain an equation for the frequency of the strategy $A$,

\begin{equation}
x(t + \epsilon) - x(t) = \epsilon\frac{x(t)[U_{A}(t) - \bar{U}(t)]}
{1-\epsilon + \epsilon \bar{U}(t)}.
\end{equation}

Now we divide both sides of (3) by $\epsilon$, perform the limit $\epsilon \rightarrow 0$,
and obtain the well known differential replicator equation:

\begin{equation}
\frac{dx(t)}{dt}=x(t)[U_{A}(t) - \bar{U}(t)].
\end{equation}
\eject 

The above equation can also be written as
\vspace{2mm}

\noindent $\frac{dx(t)}{dt}=x(t)(1-x(t))[U_{A}(t) - U_{B}(t)]$
\begin{equation}
= (a-c+d-b)x(t)(1-x(t))(x(t)-x^{*})
\end{equation}
\vspace{1mm}

For games with $m$ strategies we obtain a system of $m$ differential equations for $x_{k}(t)$,
fractions of the population playing the $k$-th strategy at the time $t$, $k=1,...,m$,

\begin{equation}
\frac{dx_{k}(t)}{dt}=x_{k}(t)[\sum_{l=1}^{m}U_{kl}x_{l}(t) - \sum_{k,l=1}^{m}U_{kl}x_{k}(t)x_{l}(t)],
\end{equation}
 
where on the right hand-size of (6) there is a difference of the average payoff of the $k$-th strategy 
and the average payoff of the population. The above system of differential equations or analogous difference
equations, called replicator dynamics was proposed in \cite{tayjon,hof1,zee}.  
For any initial condition $x^{0} \in \Delta$, it has the unique global solution, $\xi(x^{0},t)$, 
which stays in the simplex $\Delta$.

Now we review some theorems relating replicator dynamics and Nash equilibria \cite{wei,hof2,ams}.
We consider symmetric two-player games. We denote the set of strategies corresponding 
to symmetric Nash equilibria by

$$\Delta^{NE}= \{x \in \Delta: (x,x) \; is \; a \; Nash \; equilibrium\}.$$

It follows from the definition of the Nash equilibrium (see also discussion in the previous chapter 
concerning mixed strategies) that
$$\Delta^{NE}= \{x \in \Delta: u(i,x)=\max_{z \in \Delta}u(z,x) \; 
for \; every \; i \; in \; the \; support \; of \; x\}.$$

\noindent It is easy to see that 
 
$$\Delta^{0}= \{x \in \Delta: u(i,x)= u(x,x) \; for \; every \; i \; in \; the \; support \; of \; x\}$$
  
\noindent is the set of stationary points of the replicator dynamics.
\vspace{2mm}

It follows that symmetric Nash equilibria are stationary points of the replicator dynamics. 

\begin{theo}
$S \cup \Delta^{NE} \subset \Delta^{0}$
\end{theo}

The following two theorems relate stability of stationary points to Nash equilibria
\cite{wei,hof2}.

\begin{theo}
If $x \in \Delta$ is Lyapunov stable, then $x \in \Delta^{NE}.$
\end{theo} 

\begin{theo}
If $x^{0} \in interior(\Delta)$ and  $\xi(x^{0},t) \rightarrow_{t \rightarrow \infty} x$, 
then $x \in \Delta^{NE}.$
\end{theo}
\eject 
Below we present the replicator dynamics in the examples of two-player games 
discussed in the previous chapter. We write replicator equations and show their phase diagrams.
\vspace{3mm}

\noindent {\bf Stag-hunt game}

$$\frac{dx}{dt}=x(1-x)(5x-3)$$

$$\bullet<-----------\bullet------->\bullet$$
$$ \hspace{2mm} 0  \hspace{42mm}   3/5  \hspace{27mm} 1 $$
\vspace{3mm}

\noindent {\bf Hawk-Dove game}

$$\frac{dx}{dt}=-x(1-x)(x-C/V)$$

$$\bullet----------->\bullet<-------\bullet$$
$$ \hspace{1mm} 0  \hspace{41mm}   V/C \hspace{25mm} 1$$
\vspace{3mm}

\noindent {\bf Prisoner's Dilemma}

$$\frac{dx}{dt}=-x(1-x)(x+1)$$

$$\bullet<-------------------- \bullet$$
$$ \hspace{1mm} 0  \hspace{30mm}     \hspace{45mm} 1$$
\vspace{5mm}

We see that in the Stag-hunt game, both pure Nash equilibria are asymptotically stable.
The risk-dominant one has the larger basin of attraction which is true in general
because $x^{*}=(d-b)/(d-b+a-c)>1/2$ for games with an efficient equilibrium and a risk dominant one. 

In the Hawk-Dove game, the unique symmetric mixed Nash equilibrium is asymptotically stable.

In the Prisoner's Dilemma, the strategy of defection is globally asymptotically stable.

In the Rock-Scissors-Paper game, a more detailed analysis has to be done.
One can show, by straithforward computations, that the time derivative of $lnx_{1}x_{2}x_{3}$ 
is equal to zero. Therefore $lnx_{1}x_{2}x_{3}=c$ is an equation of a closed orbit for any constant $c$.
The stationary point $(1/3,1/3,1/3)$ of the replicator dynamics is Lyapunov stable and the population 
cycles on a closed trajectory (which depends on the initial condition) around its Nash equilibrium.

\section{Replicator dynamics with migration}

We discuss here a game-theoretic dynamics of a population of replicating 
who can migrate between two subpopulations or habitats \cite{migration}. 
We consider symmetric two-player games with two strategies: $A$ and $B$. 
We assume that $a>d>c$, $d>b$, and $a+b<c+d$ in a general payoff matrix given 
in the beginning of Chapter 3. Such games have two Nash equilibria: the efficient one $(A,A)$ 
in which the population is in a state with a maximal fitness (payoff) 
and the risk-dominant $(B,B)$ where players are averse to risk. 
We show that for a large range of parameters of our dynamics,
even if the initial conditions in both habitats are in the basin 
of attraction of the risk-dominant equilibrium (with respect 
to the standard replication dynamics without migration), 
in the long run most individuals play the efficient strategy.

We consider a large population of identical individuals who at each time step 
can belong to one of two different non-overlapping subpopulations or habitats 
which differ only by their replication rates. In both habitats, they take part  
in the same two-player symmetric game. Our population dynamics consists of two parts: 
the standard replicator one and a migration between subpopulations. Individuals are allowed 
to change their habitats. They move to a habitat in which the average payoff 
of their strategy is higher; they do not change their strategies.

Migration helps the population to evolve towards an efficient equilibrium.
Below we briefly describe the mechanism responsible for it. 
If in a subpopulation, the fraction of individuals playing the efficient strategy $A$ 
is above its unique mixed Nash equilibrium fraction, then the expected payoff of $A$ 
is bigger than that of $B$ in this subpopulation, and therefore the subpopulation evolves 
to the efficient equilibrium by the replicator dynamics without any migration. 
Let us assume therefore that such fraction is below the Nash equilibrium in both subpopulations. 
Without loss of generality we assume that initial conditions are such that the fraction 
of individuals playing $A$ is bigger in the first subpopulation than in the second one. 
Hence the expected payoff of $A$ is bigger in the first subpopulation than in the second one, 
and the expected payoff of $B$ is bigger in the second subpopulation than in the first one. 
This implies that a fraction of $A$-players in the second population will switch to the first one 
and at the same time a fraction of $B$-players from the first population will switch to the second one - 
migration causes the increase of the fraction of individual of the first population playing $A$. 
However, any $B$-player will have more offspring than any $A$-player 
(we are below a mixed Nash equilibrium) and this has the opposite effect on relative number 
of $A$-players in the first population than the migration. The asymptotic composition 
of the whole population depends on the competition between these two processes.

We derive sufficient conditions for migration and replication rates such that
the whole population will be in the long run in a state in which most individuals 
occupy only one habitat (the first one for the above described initial conditions) 
and play the efficient strategy.
\vspace{2mm}

Let $\epsilon$ be a time step. We allow two subpopulations 
to replicate with different speeds. We assume that during any time-step $\epsilon$, 
a fraction $\epsilon$ of the first subpopulation and a fraction $\kappa\epsilon$ 
of the second subpopulation plays the game and receives payoffs 
which are interpreted as the number of their offspring. Moreover, we allow 
a fraction of individuals to migrate to a habitat in which their strategies 
have higher expected payoffs. 

Let $r^i_s$ denote the number of individuals which use the strategy $s \in \{A, B\}$ 
in the subpopulation $i \in \{1, 2\}$. By $U^i_s$ we denote the expected payoff 
of the strategy $s$ in the subpopulation $i$:  

$$ U^1_A = ax + b(1-x), \quad   U^1_B = cx + d(1-x),$$
$$ U^2_A = ay + b(1-y), \quad   U^2_B = cy + d(1-y),$$

where
$$ x= {r^1_A\over r_1}, \quad y = {r^2_A\over r_2},  \quad r_1 = r^1_A + r^1_B,  
\quad   r_2 = r^2_A + r^2_B;$$
$x$ and $y$ denote fractions of $A$-players in the first and second population
respectively. We denote by $\alpha = {r_1\over r}$ the fraction of the whole population 
in the first subpopulation, where $r = r_1 + r_2$ is the total number of individuals.
 
The evolution of the number of individuals in each subpopulation is assumed to be a result 
of the replication and the migration flow. 
In our model, the direction and intensity of migration of individuals with a given strategy 
will be determined by the difference of the expected payoffs of that strategy in both 
habitats. Individuals will migrate to a habitat with a higher payoff. 
The evolution equations for the number of individuals playing 
the strategy $s$, $s \in \{A, B\}$, in the habitat $i$, $i \in \{1, 2\}$, have the following form: 

\begin{equation}
r^1_A(t+\epsilon) = R^1_A + \Phi_A,
\end{equation}
\begin{equation}
r^1_B(t+\epsilon) = R^1_B + \Phi_B, 
\end{equation}
\begin{equation}
r^2_A(t+\epsilon) = R^2_A - \Phi_A,
\end{equation}
\begin{equation}
r^2_B(t+\epsilon) = R^2_B - \Phi_B,
\end{equation}
where all functions on the right-hand sides are calculated at the time $t$.

Functions $R^i_s$ describe an increase of the number of the individuals playing 
the strategy $s$ in the subpopulation $i$ due to the replication: 

\begin{equation}
R^1_s= (1-\epsilon) r^1_s 
+ \delta U^1_s r^1_s,
\end{equation}
\begin{equation}
R^2_s= (1-\kappa \epsilon) r^2_s  
+ \kappa \epsilon U^2_s r^2_s,
\end{equation}

The rate of the replication of individuals playing the strategy $s$ 
in the first subpopulation is given by $\epsilon U^1_s$ , 
and in the second subpopulation by $\kappa \epsilon U^2_s$. 
The parameter $\kappa$ measures the difference of reproduction speeds 
in both habitats.   
\vspace{3mm}

Functions $\Phi_s$, $s \in \{A, B\}$, are defined by 
\begin{equation}
\Phi_s = \epsilon \gamma (U^1_s - U^2_s)[r^2_s \Theta (U^1_s - U^2_s) + r^1_s \Theta (U^2_s-U^1_s)], 
\end{equation} 

where $\Theta$ is the Heaviside's function, 
\begin{equation}
\Theta(x) = \cases {1, \ \ x\ge 0; \cr 0, \ \ x<0 \cr}
\end{equation}

and $\gamma$ is the migration rate.

\noindent Functions $\Phi_s$ describe changes of the numbers of the individuals 
playing strategy $s$ in the relevant habitat due to migration. 
$\Phi_s$ will be referred to as the migration 
of individuals (who play the strategy $s$) between two habitats. 

Thus, if for example $U^1_A > U^2_A$, then there is a migration of individuals 
with the strategy $A$ from the second habitat to the first one: 
\begin{equation}
\Phi_A = \delta \gamma r^2_A (U^1_A - U^2_A),
\end{equation}
and since then necessarily $U^1_B < U^2_B$ [note that 
$U^1_A - U^2_A = (a-b)(x-y) \hspace{3mm} and  \hspace{3mm} U^1_B - U^2_B = (c-d)(x-y)$], 
there is a migration flow of individuals with strategy $B$ from the first habitat to the second one: 
\begin{equation}
\Phi_B = \epsilon \gamma r^1_B (t) (U^1_B - U^2_B). 
\end{equation}

In this case, the migration flow $\Phi_A$ describes the increase of the number of individuals 
which play the strategy $A$ in the first subpopulation due to migration 
of the individuals playing $A$ in the second subpopulation.
This increase is assumed to be proportional to the number of individuals playing $A$ 
in the second subpopulation and the difference of payoffs of this strategy in both 
subpopulations. The constant of proportionality is $\epsilon$ times the migration rate $\gamma$. 

The case $\gamma=0$ corresponds to two separate populations which do not communicate 
and evolve independently. Our model reduces then to the standard discrete-time replicator 
dynamics. In this case, the total number of players who use a given strategy changes only due 
to the increase or decrease of the strategy fitness, as described by functions defined in (11-12). 

In the absence of the replication, there is a conservation of the number 
of individuals playing each strategy in the whole population. This corresponds to our model 
assumption that individuals can not change their strategies but only habitats in which they live. 
\vspace{2mm}

For $U^1_A > U^2_A$ we obtain from (7-10) equations for $r_{i}(t)$ and $r(t)$:
\eject

\noindent $r_{1}(t+\epsilon)=(1-\epsilon)r_{1}(t)$

\begin{equation}
+ \delta r_{1}(t)[\frac{r_{A}^{1}U_{A}^{1}+r_{B}^{1}U_{B}^{1}}{r_{1}}+
\gamma\frac{r^{2}_{A}(U_{A}^{1}-U_{A}^{2})+r^{1}_{B}(U_{B}^{1}-U_{B}^{2})}{r_{1}}],
\end{equation}

\noindent $r_{2}(t+\epsilon)=(1-\kappa \epsilon)r_{2}(t)$

\begin{equation}
+ \delta r_{2}(t)[\kappa \frac{r_{A}^{2}U_{A}^{2}+r_{B}^{2}U_{B}^{2}}{r_{2}}+
\gamma\frac{r^{2}_{A}(U_{A}^{2}-U_{A}^{1})+r^{1}_{B}(U_{B}^{2}-U_{B}^{1})}{r_{2}}],
\end{equation}

\noindent $r(t+\delta)= (1-\epsilon)r_{1}(t)+ (1-\kappa \delta)r_{2}(t)$

\begin{equation} 
+ \delta r(t)[\alpha ({\frac{r^1_A}{r_1}}U_{A}^{1}+ {\frac{r^1_B}{r_1}} U_{B}^{1})+
(1-\alpha)\kappa ( {\frac{r^2_A}{r_2}} U_{A}^{2}+ {\frac{r^2_B}{r_2}} U_{B}^{2})],
\end{equation}

where all functions in square brackets depend on $t$. 
\vspace{2mm}

Now, like in the derivation of the standard replicator dynamics, we consider frequencies 
of individuals playing the relevant strategies in both habitats. 
Thus, we focus on the temporal evolution of the frequencies, $x$ and $y$, and the relative 
size of the first subpopulation, $\alpha$. We divide (7) by (17), (9) by (18), and (17) by (19).
Performing the limit $\epsilon \rightarrow 0$ we obtain the following differential equations:
\vspace{2mm}

\noindent $\frac{dx}{dt}=x[(1-x)(U_{A}^{1}-U_{B}^{1})$

\begin{equation}
+ \gamma[(\frac{y(1-\alpha)}{x\alpha}-\frac{y(1-\alpha)}{\alpha})
(U_{A}^{1}-U_{A}^{2})-(1-x)(U_{B}^{1}-U_{B}^{2})]],
\end{equation}

\begin{equation}
\frac{dy}{dt}=y[\kappa(1-y)(U_{A}^{2}-U_{B}^{2})+\gamma[(1-y)(U_{A}^{2}-U_{A}^{1})-
\frac{(1-x)\alpha}{1-\alpha}(U_{B}^{2}-U_{B}^{1})]],
\end{equation}

\noindent $\frac{d\alpha}{dt}=\alpha(1-\alpha)[xU_{A}^{1}+(1-x)U_{B}^{1}-(yU_{A}^{2}+(1-y)U_{B}^{2})]$
\vspace{2mm}

\noindent \hspace{10mm} $+ \alpha \gamma[\frac{y(1-\alpha)}{\alpha}(U_{A}^{1}-U_{A}^{2})+(1-x)(U_{B}^{1}-U_{B}^{2})]$
\begin{equation}
+ \alpha(1-\alpha)(\kappa-1)(1-yU^{2}_{A}-(1-y)U^{2}_{B}).
\end{equation}
  
Similar equations are derived for the case $U^1_A < U^2_A$ (since our model is symmetric 
with respect to the permutation of the subpopulations, it is enough to renumerate the 
relevant indices and redefine the parameter $\kappa$). 

Assume first that $U_{A}^{1}(0)>U_{A}^{2}(0)$, which is equivalent to  
$x(0)>y(0)$. It follows from (7-10) that a fraction of $A$-players from the subpopulation $2$
will migrate to the subpopulation $1$ and a fraction of $B$-players will migrate in the opposite direction.
This will cause $x$ to increase and $y$ to decrease. However, if $x(0)<x^{*}$ and $y(0)<x^{*}$, 
then $U_{A}^{1}<U_{B}^{1}$ and $U_{A}^{2}<U_{B}^{2}$, therefore $B$-players will have more 
offspring than $A$-players. This has the opposite effect on the relative number of $A$-players 
in the first subpopulation than migration. If $x(0)<y(0)$, then migration takes place 
in the reverse directions.

The outcome of the competition between migration and replication depends, for a given payoff matrix, 
on the relation between $x(0)-y(0)$, $\gamma$ and $\kappa$. 
We are interested in formulating sufficient conditions for the parameters of the model, for which 
most individuals of the whole population will play in the long run the efficient strategy $A$. 
We prove the following theorem \cite{migration}.

\begin{theo}

If 
$$\gamma [ x(0)-y(0) ] > max [\frac {d-b} {d-c}, \frac {\kappa (a-c)}{a-b}],$$
then $x(t) \rightarrow_{t \rightarrow \infty}1$ and $y(t) \rightarrow_{t \rightarrow \infty} 0$.

If $\kappa<(a-1)/(d-1)$, then $\alpha(t) \rightarrow_{t \rightarrow \infty} 1$. 
\vskip 0.2cm

If 
$$\gamma [ y(0)-x(0) ] > max [\frac {\kappa (d-b)} {d-c}, \frac {a-c} {a-b}],$$
then $x(t) \rightarrow_{t \rightarrow \infty}0$ and $y(t) \rightarrow_{t \rightarrow \infty} 1$.

If $\kappa > (d-1)/(a-1)$, then $\alpha(t) \rightarrow_{t \rightarrow \infty} 0.$ 
\end{theo}

{\bf Proof}: 

Assume first that $x(0)>y(0)$. From (20-21) we get the following differential inequalities:

\begin{equation}
\frac{dx}{dt} > x(1-x)[U_{A}^{1}-U_{B}^{1})+\gamma(U_{B}^{2}-U_{B}^{1})],
\end{equation}

\begin{equation}
\frac{dy}{dt} < y(1-y)[\kappa(U_{A}^{2}-U_{B}^{2})+\gamma(U_{A}^{2}-U_{A}^{1})],
\end{equation}

Using explicit expressions for $U_{s}^{i}$ we get

\begin{equation}
\frac{dx}{dt} > x(1-x)[(a-c+d-b)x+b-d+\gamma(d-c)(x-y)], 
\end{equation}

\begin{equation}
\frac{dy}{dt} < y(1-y)[\kappa[(a-c+d-b)y+b-d]-\gamma(a-b)(x-y)],
\end{equation}

We note that if $\gamma(d-c)(x(0)-y(0)) > d - b$  
then $\gamma(d-c)(x(0)-y(0))+ b - d + (a-c+d-b)x(0) > 0,$ i.e. $dx/dt(0)>0$.  

Analogously, if $\gamma(a-b)(x(0)-y(0))> \kappa (a-c)$, then 
$\gamma(a-b)(x(0)-y(0))> \kappa [(a-c+d-b) + b-d] > \kappa [(a-c+d-b)y(o) + b-d],$ 
therefore $dy/dt(0)<0$. 
Thus, combining both conditions we conclude that $x(t)-y(t)$ is an increasing function so
$x(t)>y(t)$ for all $t \geq 0$, hence we may use (20-22) all the time. 
We get that
$x(t) \rightarrow_{t \rightarrow \infty}1$ and $y(t) \rightarrow_{t \rightarrow \infty} 0$, 
and the first part of the thesis follows. 
Now from (22) it follows that if $a-d+(\kappa-1)(1-d)>0$, i.e. $\kappa< (a-1)/(d-1)$, 
then $\alpha(t) \rightarrow_{t \rightarrow \infty} 1$. 

The second part of Theorem 5, corresponding to initial conditions $y(0)>x(0),$ 
can be proved analogously, starting from eqs. (7-10) written for the case $U^1_A(0) < U^2_A(0)$ 
and their continuous counterparts. We omit details. 
\vspace{2mm}

The above conditions for $\kappa$ mean that the population consisting of just $A$-players 
replicates faster (exponentially in $(a-1)t$) than the one consisting of just $B$-players 
(exponentially in $(d-1)\kappa t$). The same results would follow if the coefficients 
of the payoff matrix of the game played in one habitat would differ from those 
in the second habitat by an additive constant.
 
We showed that introduction of the  mechanism of attraction by the habitat 
with a higher expected payoff in the standard replicator dynamics helps 
the whole population to reach the state in which in the long run most individuals 
play the efficient strategy. 

More precisely, we proved that for a given rate of migration, if the fractions of individuals 
playing the efficient strategy in both habitats are not too close to each other, then the habitat 
with a higher fraction of such players overcomes the other one in the long run. 
The fraction of individuals playing the efficient strategy tends to unity in this habitat 
and consequently in the whole population. Alternatively, we may say that the bigger 
the rate of migration is, larger is the basin of attraction of the efficient equilibrium. 
In particular, we showed that for a large range of parameters of our dynamics,
even if the initial conditions in both habitats are in the basin 
of attraction of the risk-dominant equilibrium (with respect 
to the standard replication dynamics without migration), 
in the long run most individuals play the efficient strategy.

\section{Replicator dynamics with time delay}

Here we consider two-player games with two strategies, two pure non-symmetric
Nash equilibria, and a unique symmetric mixed one, that is $a<c$ and $d<b$ 
in a general payoff matrix given in the beginning of Chapter 3.
Let us recall that the Hawk-Dove game is of such type.  

Recently Tao and Wang \cite{taowang} investigated the effect of a time delay 
on the stability of the mixed equilibrium in the replicator dynamics. 
They showed that it is asymptotically stable if a time delay is small. 
For sufficiently large delays it becomes unstable. 

We construct two models of discrete-time replicator dynamics
with a time delay \cite{delay}. In the social-type model, players imitate opponents 
taking into account average payoffs of games played some units of time ago.
In the biological-type model, new players are born from parents who played
in the past. We show that in the first type of dynamics, the unique symmetric mixed Nash equilibrium
is asymptotically stable for small time delays and becomes unstable for large ones
when the population oscillates around its stationary state. 
In the second type of dynamics, however, the Nash equilibrium 
is asymptotically stable for any time delay. Our proofs are elementary, 
they do not rely on the general theory of delay differential and difference equations.

\subsection{Social-type time delay}

\noindent Here we assume that individuals at time $t$ replicate due to average payoffs 
obtained by their strategies at time $t-\tau$ for some delay $\tau>0$
(see also a discussion after (32)). 
As in the standard replicator dynamics, we assume that during the small time interval $\epsilon$, 
only an $\epsilon$ fraction of the population takes part in pairwise competitions, that is plays games.
Let $r_{i}(t)$, $i=A, B,$ be the number of individuals playing at the time $t$ the strategy 
$A$ and $B$ respectively, $r(t)=r_{A}(t)+r_{B}(t)$ the total number of players and
$x(t)=\frac{r_{1}(t)}{r(t)}$ a fraction of the population playing $A$.

We propose the following equations:

\begin{equation}
r_{i}(t + \epsilon) = (1-\epsilon)r_{i}(t) + \epsilon r_{i}(t)U_{i}(t-\tau); \; \; i= A,B.
\end{equation}

Then for the total number of players we get
\begin{equation}
r(t + \epsilon) = (1-\epsilon)r(t) + \epsilon r(t)\bar{U}_{o}(t-\tau),
\end{equation}

where $\bar{U}_{o}(t-\tau)=x(t)U_{A}(t-\tau)+(1-x(t))U_{B}(t-\tau).$
\vspace{2mm}

\noindent We divide (27) by (28) and obtain an equation for the frequency of the strategy $A$,

\begin{equation}
x(t + \epsilon) - x(t) = \epsilon \frac{x(t)[U_{A}(t-\tau) - \bar{U}_{o}(t-\tau)]}
{1-\epsilon + \epsilon \bar{U}_{o}(t-\tau)} 
\end{equation}

and after some rearrangements we get

\begin{equation}
x(t + \epsilon) - x(t) = -\epsilon x(t)(1-x(t))[x(t-\tau)-x^{*}]\frac{\delta}{1-\epsilon + 
\epsilon \bar{U}_{o}(t-\tau)}, 
\end{equation}

where $x^{*}= (d-b)/(d-b+a-c)$ is the unique mixed Nash equilibrium of the game.
\vspace{2mm}

Now the corresponding replicator dynamics in the continuous time reads

\begin{equation}
\frac{dx(t)}{dt}=x(t)[U_{A}(t-\tau) - \bar{U}_{o}(t-\tau)]
\end{equation}

and can also be written as
\vspace{2mm}

\noindent $\frac{dx(t)}{dt}=x(t)(1-x(t))[U_{A}(t-\tau) - U_{B}(t-\tau)]$

\begin{equation}
= -\delta x(t)(1-x(t))(x(t-\tau)-x^{*}).
\end{equation}

The first equation in (32) can be also interpreted as follows. Assume that randomly chosen players 
imitate randomly chosen opponents. Then the probability that a player who played $A$ 
would imitate the opponent who played $B$ at time $t$ is exactly $x(t)(1-x(t)).$ 
The intensity of imitation depends on the delayed information about the difference 
of corresponding payoffs at time $t- \tau$. We will therefore say that such models 
have a social-type time delay.

Equations (31-32) are exactly the time-delay replicator dynamics proposed and analyzed 
by Tao and Wang \cite{taowang}. They showed that if $\tau< c-a+b-d \pi /2(c-a)(b-d)$,
then the mixed Nash equilibrium, $x^{*}$, is asymptotically stable. 
When $\tau$ increases beyond the bifurcation value $c-a+b-d \pi /2(c-a)(b-d)$, $x^{*}$ 
becomes unstable. We have the following theorem \cite{delay}.

\begin{theo}
$x^{*}$ is asymptotically stable in the dynamics (30) if $\tau$ is sufficiently small
and unstable for large enough $\tau$.
\end{theo}

\noindent {\bf Proof:} We will assume that $\tau$ is a multiple of $\epsilon$, 
$\tau=m\epsilon$ for some natural number $m$. Observe first that if $x(t - \tau) < x^{*}$, 
then $x(t + \epsilon) > x(t)$, and if $x(t - \tau) > x^{*}$, 
then $x(t + \epsilon) < x(t)$. Let us assume first that there is $t'$ 
such that $x(t'), x(t'-\epsilon), x(t'-2\epsilon),..., x(t'-\tau) < x^{*}$.
Then $x(t)$, $t \geq t'$ increases up to the moment $t_{1}$ for which
$x(t_{1} - \tau) > x^{*}$. If such $t_{1}$ does not exist then 
$x(t) \rightarrow_{t \rightarrow \infty}x^{*}$ and the theorem is proved.
Now we have $x^{*}<x(t_{1}-\tau) < x(t_{1}-\tau+\epsilon)< \ldots <x(t_{1})$
and $x(t_{1}+\epsilon)<x(t_{1})$ so $t_{1}$ is a turning point. Now $x(t)$
decreases up to the moment $t_{2}$ for which $x(t_{2}-\tau) < x^{*}$.
Again, if such $t_{2}$ does not exist, then the theorem follows.
Therefore let us assume that there is an infinite sequence, $t_{i}$, of such turning points.
Let $\eta_i = |x(t_{i}) - x^{*}|$. We will show that 
$\eta_i \rightarrow_{i \rightarrow \infty} 0$.
  
For $t \in \{t_{i}, t_{i} + \epsilon, \ldots, t_{i + 1}-1\}$ we have the following bound 
for $x(t + \epsilon) - x(t)$:
 
\begin{equation}
|x(t + \epsilon) - x(t)| < \frac{1}{4} \eta_{i} \frac{\epsilon \delta}
{1 -\epsilon + \epsilon \bar{U}_{o}(t - \tau)}.
\end{equation}
  
This means that

\begin{equation}
\eta_{i+1} <  (m+1)\epsilon K \eta_{i},
\end{equation}  

where $K$ is the maximal possible value of 
$\frac{\delta}{4(1-\epsilon + \epsilon \bar{U}_{o}(t - \tau))}.$

We get that if 

\begin{equation}
\tau < \frac{1}{K}-\epsilon,
\end{equation}    

then $\eta_{i}\rightarrow_{i \rightarrow \infty} 0$ so $x(t)$ converges to $x^{*}$. 
\vspace{2mm}

Now if for every $t$, 
$|x(t + \epsilon)-x^{*}| < \max_{k\in \{0,1,...,m\}}|x(t - k\epsilon)-x^{*}|$, 
then $x(t)$ converges to $x^{*}$. 
Therefore assume that there is $t''$ such that
$|x(t'' + \epsilon)-x^{*}| \geq \max_{k\in \{0,1,...,m\}}|x(t'' - k\epsilon)-x^{*}|$. 
If $\tau$ satisfies (35), then it follows that 
$x(t+\epsilon),...,x(t+\epsilon+\tau)$ are all on the same side of $x^{*}$ 
and the first part of the proof can be applied. We showed that $x(t)$ converges to $x^{*}$ 
for any initial conditions different from $0$ and $1$ hence 
$x^{*}$ is globally asymptotically stable.

Now we will show that $x^{*}$ is unstable for any large enough $\tau$.

Let $\gamma>0$ be arbitrarily small and consider a following perturbation 
of the stationary point $x^{*}$: $x(t)=x^{*}, t\leq 0$ 
and $x(\epsilon) = x^{*}+\gamma$. It folows from (30) 
that $x(k\epsilon)=x(\epsilon)$ for $k=1,...,m+1$. 
Let $K'=\min_{x \in[x^{*}-\gamma,x^{*}+\gamma]}
\frac{x(1-x)\delta}{4(1-\epsilon + \epsilon \bar{U}_{o}(t - \tau))}$. 
If $\frac{m}{2}\epsilon K'\gamma > 2\gamma$, that is 
$\tau > \frac{4}{K'}$, then it follows from (30) 
that after $m/2$ steps (we assume without loss of generality that $m$ is even)
$x((m+1+m/2)\epsilon)<x^{*}-\gamma$. In fact we have 
$x((2m+1)\epsilon)< \ldots < x((m+1)\epsilon)$ and at least $m/2$ of $x's$ in this sequence
are smaller than $x^{*}-\gamma$. Let $\bar{t}>(2m+1)\epsilon$ be the smallest 
$t$ such that $x(t)>x^{*}-\gamma$. Then we have 
$x(\bar{t}-m \epsilon), \ldots, x(\bar{t}-\epsilon) < x^{*}-\gamma < x(\bar{t})$
hence after $m/2$ steps, $x(t)$ crosses $x^{*}+\gamma$ and the situation repeats itself.
\vspace{2mm}

\noindent We showed that if
\begin{equation}
\tau >\frac{4}{K'}, 
\end{equation}   
then there exists an infinite sequence, $\tilde{t}_{i}$, such that $|x(\tilde{t}_{i})-x^{*}|>\gamma$
and therefore $x^{*}$ is unstable. Moreover, $x(t)$ oscillates around $x^{*}$.

\subsection{Biological-type time delay}

\noindent Here we assume that individuals born at time $t-\tau$ are able to take part in contests 
when they become mature at time $t$ or equivalently they are born $\tau$ units of time 
after their parents played and received payoffs. We propose the following equations:

\begin{equation}
r_{i}(t + \epsilon) = (1-\epsilon)r_{i}(t) + \epsilon r_{i}(t-\tau)U_{i}(t-\tau); \; \; i= A,B.
\end{equation}
Then the equation for the total number of players reads
\vspace{2mm}

\noindent $r(t + \epsilon) = (1-\epsilon)r(t)$ 

\begin{equation}
+ \epsilon r(t)[\frac{x(t)r_{A}(t-\tau)}{r_{A}(t)}U_{A}(t-\tau)
+\frac{(1-x(t))r_{B}(t-\tau)}{r_{B}(t)}U_{B}(t-\tau)].
\end{equation}

\noindent We divide (37) by (38) and obtain an equation for the frequency of the first strategy,

\begin{equation}
x(t + \epsilon) - x(t) = \epsilon \frac{x(t - \tau)U_{A}(t - \tau) - x(t)\bar{U}(t - \tau)}
{(1-\epsilon)\frac{r(t)}{r(t-\tau)} + \epsilon \bar{U}(t-\tau)},
\end{equation}

\noindent where $\bar{U}(t-\tau)=x(t-\tau)U_{A}(t-\tau)+(1-x(t-\tau))U_{B}(t-\tau).$
\vspace{2mm}

We proved in \cite{delay} the following 

\begin{theo}
$x^{*}$ is asymptotically stable in the dynamics (39) for any value of the time delay $\tau$. 
\end{theo}

We begin by showing our result in the following simple example.
\vspace{3mm}

\noindent The payoff matrix is given by  
$U = \left(\begin{array}{cc}
    0 & 1\\
    1 & 0
  \end{array}\right)$ hence $x^{*} = \frac{1}{2}$ is the mixed Nash equilibrium
which is asymptotically stable in the replicator dynamics without the time delay.
The equation (39) now reads

\begin{equation}
x(t+\epsilon)-x(t)=\epsilon \frac{x(t-\tau)(1 - x(t-\tau))-2 x(t)x(t-\tau)(1-x(t-\tau))}
{(1-\epsilon)\frac{r(t)}{r(t-\tau)}+2\epsilon x(t-\tau)(1-x(t-\tau))} 
\end{equation}

After simple algebra we get
\vspace{2mm}

\noindent $x(t+ \epsilon) - \frac{1}{2} + \frac{1}{2} - x(t)$

\begin{equation}
= \epsilon (1 - 2x(t)) \frac{x(t-\tau)(1-x(t - \tau))}
{(1-\epsilon)\frac{r(t)}{r(t - \tau)} + 2\epsilon x(t - \tau)(1 - x (t - \tau))},
\end{equation}

so 
$$x(t + \epsilon) - \frac{1}{2} = (x(t)- \frac{1}{2}) \frac{1}
{1 + \frac{\epsilon r(t - \tau)}{(1-\epsilon)r(t)}2x(t - \tau)(1- x(t - \tau))}$$ 

hence

\begin{equation}
|x(t+\epsilon)-\frac{1}{2}| < |x(t)-\frac{1}{2}|. 
\end{equation}
  
It follows that $x^{*}$ is globally asymptotically stable.
\vspace{4mm}

Now we present the proof for the general payoff matrix with a unique symmetric mixed Nash equilibrium.
\vspace{3mm}

\noindent {\bf Proof of Theorem 7:} 
\vspace{3mm}

Let $c_{t} = \frac{x(t)U_{A}(t)}{\bar{U}(t)}$. 
Observe that if $x(t) < x^{*}$, then $c_{t} > x(t)$, if $x(t) > x^{*}$, 
then $c_{t} < x(t)$, and if $x(t) = x^{*}$, then $c_{t} = x^{*}$. 
We can write (39) as
\vspace{2mm}

\noindent $x(t + \epsilon) - x(t)$

\begin{equation}
= \epsilon\frac{x(t - \tau)U_{A}(t - \tau) - c_{t - \tau}
\bar{U}(t - \tau) + c_{t - \tau} \bar{U}(t - \tau) - x(t) \bar{U}(t - \tau)}
{(1-\epsilon)\frac{p(t)}{p(t - \tau)} + \epsilon \bar{U}(t - \tau)}
\end{equation}
and after some rearrangements we obtain

\begin{equation}
x(t + \epsilon) - c_{t - \tau} = (x(t) - c_{t - \tau}) \frac{1}
{1 + \frac{\epsilon p(t - \tau)}{(1-\epsilon) p(t)} \bar{U}(t - \tau)}. 
\end{equation}

We get that at time $t + \epsilon$, $x$ is closer to $c_{t -\tau}$ than at time $t$
and it is on the same side of $c_{t -\tau}$.
We will show that  $c$ is an increasing or a constant function of $x$. 
Let us calculate the derivative of $c$ with respect to $x$. 

\begin{equation}
c' = \frac{f(x)}{(xU_{A} + (1 - x)U_{B})^2}, 
\end{equation}
where  
\begin{equation}
f(x)= (ac + bd - 2ad)x^{2}+ 2d(a-b)x+bd.
\end{equation}

A simple analysis shows that $f>0$ on $(0,1)$ or $f=0$ on $(0,1)$
(in the case of $a=d=0$). Hence $c(x)$ is either an increasing
or a constant function of $x$. In the latter case, $\forall_x c(x) = x^{*},$ 
as it happens in our example, and the theorem follows.
  
We will now show that
\begin{equation}
|x(t + \tau + \epsilon) - x^{*}| < \max\{|x(t) - x^{*}|, |x(t + \tau) - x^{*}|\} 
\end{equation}
hence $x(t)$ converges to $x^{*}$ for any initial conditions different from
$0$ and $1$ so $x^{*}$ is globally asymptotically stable.
\vspace{2mm}

If $x(t) < x^{*}$ and $x(t + \tau) <  x^{*}$, then $x(t) < c_{t} \leq x^{*}$ 
and also $x(t+\tau) < c_{t+\tau} \leq x^{*}$. 

From (44) we obtain

  \[ \left\{\begin{array}{cc}
       x \left( t + \tau \right) < x \left( t + \tau + \epsilon \right) < c_t 
       \; \; if \; \;  x \left( t + \tau \right) < c_t\\
       x \left( t \right) < x \left( t + \tau + \epsilon \right) = c_t  \; \; if \; \; 
       x \left( t + \tau \right) = c_t\\
       x \left( t \right) < c_t < x \left( t + \tau + \epsilon \right) < x \left( t +
       \tau \right)\; \;  if \; \;  x \left( t + \tau \right) > c_t
     \end{array}\right. \]
hence (47) holds.

If $x \left( t \right) > x^{\ast}$ and $x \left( t + \tau \right) <
  x^{\ast}$, then $x \left( t + \tau \right) < x^{\ast} < c_t < x \left( t
  \right)$ and either $x \left( t + \tau \right) <
  x \left( t + \tau + \epsilon \right) < x^{\ast}$ or $x^{\ast} < x \left( t + \tau
  + \epsilon \right) < c_t$ which means that (47) holds.
  
The cases of $x \left( t \right) > x^{*}$, $x \left( t + \tau \right) >
  x^{*}$ and $x \left( t \right) < x^{*}$, $x \left( t + \tau \right)<x^{*}$
can be treated analogously. We showed that (47) holds.

\section{Stochastic dynamics of finite populations}

In the next two chapters we will discuss various stochastic dynamics of populations with a fixed number 
of players interacting in discrete moments of time. We will analyze symmetric two-player games 
with two or three strategies and multiple Nash equilibria. We will address the problem 
of equilibrium selection - which strategy will be played in the long run with a high frequency. 

Our populations are characterized either by numbers of individuals playing respective strategies
in well-mixed populations or by a complete profile - assignment of strategies to players 
in spatial games. Let $\Omega$ be a state space of our system. For non-spatial games 
with two strategies, $\Omega=\{0,1,...,n\}$, where $n$ is the number of players 
or $\Omega= 2^{\Lambda}$ for spatial games with players located on the finite subset 
$\Lambda$ of ${\bf Z}, {\bf Z}^{2}$, or any other infinite graph, and interacting 
with their neighbours.  In well-mixed populations, in discrete moments of times, 
some individuals switch to a strategy with a higher mean payoff. 
In spatial games, players choose strategies which are best responses, i.e. 
ones which maximize the sum of the payoffs obtained from individual games. 
The above rules define deterministic dynamics with some stochastic part 
corresponding to a random matching of players or a random choice of players who may revise their strategies.
We call this mutation-free or noise-free dynamics. It is a Markov chain with a state space $\Omega$ 
and a transition matrix $P^{0}$. We are especially interested in absorbing states, i.e. rest points
of our mutation-free dynamics. Now, with a small probability, $\epsilon$, players may mutate
or make mistakes of not chosing the best reply. The presence of mutatation allows the system 
to make a transition from any state to any other state with a positive probability in some finite number 
of steps or to stay indefinitively at any state for an arbitrarily long time. 
This makes our Markov chains with a transition matrix $P^{\epsilon}$ ergodic ones. They have therefore
unique stationary measures. To describe the long-run behavior of stochastic dynamics of finite populations,
Foster and Young \cite{foya} introduced a concept of stochastic stability. A state of the system is {\bf stochastically stable} if it has a positive probability in the stationary measure of the corresponding Markov chain 
in the zero-noise limit, that is the zero probability of mistakes or the zero-mutation level. 
It means that along almost any time trajectory the frequency of visiting this state converges to a positive value 
given by the stationary measure. Let $\mu^{\epsilon}$ be the stationary measure of our Markov chain.  

\begin{defi}
$X \in \Omega$ is {\bf stochastically stable} 
if $\lim_{\epsilon \rightarrow 0}\mu^{\epsilon}(X) > 0.$
\end{defi} 

It is a fundamental problem to find stochastically stable states for any stochastic dynamics of interest.
We will use the following tree representation of stationary measures of Markov chains 
proposed by Freidlin and Wentzell \cite{freiwen1,freiwen2}, see also \cite{shub}.
Let $(\Omega,P^{\epsilon})$ be an ergodic Markov chain with a state space 
$\Omega$, transition probabilities given by the transition matrix 
$P^{\epsilon}: \Omega \times \Omega \rightarrow [0,1]$, 
where $P^{\epsilon}(Y,Y')$ is a conditional probability 
that the system will be in the state $Y' \in \Omega$ at the time $t+1$, 
if it was in the state $Y \in \Omega$ at the time $t$, 
and a unique stationary measure, $\mu^{\epsilon}$, also called a stationary state. 
A stationary state is an eigenvector of $P^{\epsilon}$ corresponding to the eigenvalue $1$,
i.e. a solution of a system of linear equations,

\begin{equation}
\mu^{\epsilon}P^{\epsilon}=\mu^{\epsilon},
\end{equation}
where $\mu^{\epsilon}$ is a row wector $[\mu^{\epsilon}_{1},...,\mu^{\epsilon}_{|\Omega|}]$.
After specific rearrangements one can arrive at an expression for the stationary state 
which involves only positive terms. This will be very useful in describing the asymptotic 
behaviour of stationary states. 

For $X \in \Omega$, let an X-tree be a directed graph on $\Omega$ such that from every $Y \neq X$ 
there is a unique path to $X$ and there are no outcoming edges out of $X$. 
Denote by $T(X)$ the set of all X-trees and let 
\begin{equation}
q^{\epsilon}(X)=\sum_{d \in T(X)} \prod_{(Y,Y') \in d}P^{\epsilon}(Y,Y'),
\end{equation}
where the product is with respect to all edges of $d$. 
\vspace{2mm}

We have that
\begin{equation}
\mu^{\epsilon}(X)=\frac{q^{\epsilon}(X)}{\sum_{Y \in \Omega}q^{\epsilon}(Y)}
\end{equation}

for all $X \in \Omega.$
\vspace{1mm}

We assume that our noise-free dynamics, i.e. in the case of  $\epsilon=0$, has at least one absorbing state 
and there are no absorbing sets (recurrent classes) consisting of more than one state. 
It then follows from (50) that only absorbing states can be stochastically stable.

Let us begin with the case of two absorbing states, $X$ and $Y$. Consider a dynamics in which 
$P^{\epsilon}(Z,W)$ for all $Z, W \in \Omega$, is of order $\epsilon^{m}$, 
where $m$ is the number of mistakes involved to pass from $Z$ to $W$. The noise-free limit 
of $\mu^{\epsilon}$ in the form (50) has a $0/0$ character. Let $m_{XY}$ be a minimal number 
of mistakes needed to make a transition from the state $X$ to $Y$ and $m_{YX}$ the minimal number of mistakes
to evolve from $Y$ to $X$. Then $q^{\epsilon}(X)$ is of the order $\epsilon^{m(YX)}$ and $q^{\epsilon}(Y)$ is of the order $\epsilon^{m(XY)}$. If for example $m_{YX} < m_{XY}$, then $ \lim_{\epsilon \rightarrow 0}\mu^{\epsilon}(X)=1$ 
hence $X$ is stochastically stable.   

In general, to study the zero-noise limit of the stationary measure, 
it is enough to consider paths between absorbing states. More precisely, 
we construct X-trees with absorbing states $X^{k}$, $k=1,...,l$ as vertices; 
the family of such $X$-trees is denoted by $\tilde{T}(X)$. Let 
\begin{equation}
q_{m}(X)=max_{d \in \tilde{T}(X)} \prod_{(Y,Y') \in d}\tilde{P}(Y,Y'),
\end{equation}
where $\tilde{P}(Y,Y')= max \prod_{(W,W')}P(W,W')$,
where the product is taken along any path joining $Y$ with $Y'$ and the maximum 
is taken with respect to all such paths. 
Now we may observe that if $lim_{\epsilon \rightarrow 0} q_{m}(X^{i})/q_{m}(X^{k})=0,$
for every $i=1,...,l$, $i \neq k$, then $X^{k}$ is stochastically stable. 
Therefore we have to compare trees with the biggest products in (51); 
such trees are called maximal.

The above characterisation of the stationary measure
was used to find stochastically stable states in non-spatial \cite{kmr,young1,rvr,vr,young2,mar}
and spatial games \cite{ellis1,ellis2}. We will use it below in our examples. 

In many cases, there exists a state $X$ such that $\lim_{\epsilon \rightarrow 0}\mu^{\epsilon}(X)=1$ 
in the zero-noise limit. Then we say that $X$ was selected in the zero-noise limit of a given stochastic dynamics. 
However, for any low but fixed mutation level, when the number of players is very large, 
the frequency of visiting any single state can be arbitrarily low. 
It is an ensemble of states that can have a probability close to one in the stationary measure.
The concept of the ensemble stability is discussed in Chapter 9.

\section{Stochastic dynamics of well-mixed populations}

Here we will discuss stochastic dynamics of well-mixed populations of players 
interacting in discrete moments of time. We will analyze two-player games 
with two strategies and two pure Nash equilibria. The efficient strategy
(also called payoff dominant) when played by the whole population results 
in its highest possible payoff (fitness). The risk-dominant one is played by individuals
averse to risk. The strategy is risk dominant if it has a higher expected payoff against 
a player playing both strategies with equal probabilities \cite{hs}. We will address the problem 
of equilibrium selection - which strategy will be played in the long run with a high frequency.

We will review two models of dynamics of a population with a fixed number 
of individuals. In both of them, the selection part of the dynamics ensures that if 
the mean payoff of a given strategy is bigger than the mean payoff of the other one, 
then the number of individuals playing the given strategy increases. 
In the first model, introduced by Kandori, Mailath, and Rob \cite{kmr}, 
one assumes (as in the standard replicator dynamics) that individuals receive average payoffs 
with respect to all possible opponents - they play against the average strategy. 
In the second model, introduced by Robson and Vega-Redondo \cite{rvr}, at any moment of time, 
individuals play only one or few games with randomly chosen opponents.
In both models, players may mutate with a small probability, hence the population may move 
against a selection pressure. Kandori, Mailath, and Rob showed that in their model, 
the risk-dominant strategy is stochastically stable - if the mutation level is small enough 
we observe it in the long run with the frequency close to one \cite{kmr}. 
In the model of Robson and Vega-Redondo, the efficient strategy is stochastically stable \cite{rvr,vr}. 
It is one of very few models in which an efficient strategy 
is stochastically stable in the presence of a risk-dominant one. 
The population evolves in the long run to a state with the maximal fitness. 

The main goal of this chapter is to investigate the effect of the number of players
on the long-run behaviour of the Robson-Vega-Redondo model \cite{population}. 
We will discuss a sequential dynamics and the one where each individual enjoys 
each period a revision opportunity with the same probability.   
We will show that for any arbitrarily low but a fixed level of mutations, 
if the number of players is sufficiently large, then a risk-dominant strategy is played 
in the long run with a frequency closed to one -  
a stochastically stable efficient strategy is observed with a very low frequency. 
It means that when the number of players increases, the population undergoes a transition 
between an efficient payoff-dominant equilibrium and a risk-dominant one. 
We will also show that for some range of payoff parameters, stochastic stability itself 
depends on the number of players. If the number of players is below certain value 
(which may be arbitrarily large), then a risk-dominant strategy is stochastically stable. 
Only if the number of players is large enough, an efficient strategy becomes stochastically stable as proved 
by Robson and Vega-Redondo. 

Combining the above results we see that for a low but fixed 
noise level, the population undergoes twice a transition between its two equilibria 
as the number of individuals increases \cite{banach}. In addition, for a sufficiently 
large number of individuals, the population undergoes another equilibrium transition 
when the noise decreases.  

Let us formally introduce our models. We will consider a finite population of $n$ individuals
who have at their disposal one of two strategies:
$A$ and $B$. At every discrete moment of time, $t=1,2,...$
individuals are randomly paired (we assume that $n$ is even)
to play a two-player symmetric game with payoffs given by the following matrix:
\vspace{3mm}

\hspace{22mm} A  \hspace{5mm} B   
\vspace{1mm}

\hspace{15mm} A \hspace{3mm} a  \hspace{5mm} b 

U = \hspace{6mm} 

\hspace{15mm} B \hspace{3mm} c  \hspace{5mm} d,
\vspace{3mm}

where $a>c, d>b, a>d$, and $a+b<c+d$ so $(A,A)$ is an efficient Nash equilibrium 
and $(B,B)$ is a risk-dominant one. 
\vspace{2mm}

At the time $t$, the state of our population is described by the number of individuals, $z_{t}$, playing $A$. 
Formally, by the state space we mean the set $$\Omega=\{z, 0 \leq z\leq n\}.$$
Now we will describe the dynamics of our system.
It consists of two components: selection and mutation.
The selection mechanism ensures that if the mean payoff 
of a given strategy, $\pi_{i}(z_{t}), i=A,B$, 
at the time $t$ is bigger than the mean payoff 
of the other one, then the number of individuals
playing the given strategy increases in $t+1$. 
In their paper, Kandori, Mailath, and Rob \cite{kmr} write
\begin{equation}
\pi_{A}(z_{t})=\frac{a(z_{t}-1)+b(n-z_{t})}{n-1},
\end{equation} 
$$\pi_{B}(z_{t})=\frac{cz_{t}+d(n-z_{t}-1)}{n-1},$$
provided $0<z_{t}<n$.

It means that in every time step, players are paired infnitely many times
to play the game or equivalently, each player plays with every other player 
and his payoff is the sum of corresponding payoffs. This model may be therefore
considered as an analog of replicator dynamics for populations with a fixed numbers 
of players.

The selection dynamics is formalized in the following way:

\begin{equation}
z_{t+1} > z_{t} \hspace{2mm} if \hspace{2mm} \pi_{A}(z_{t}) > \pi_{B}(z_{t}),
\end{equation}
$$z_{t+1} < z_{t} \hspace{2mm} if \hspace{2mm} \pi_{A}(z_{t}) < \pi_{B}(z_{t}),$$
$$z_{t+1}= z_{t} \hspace{2mm} if \hspace{2mm} \pi_{A}(z_{t}) = \pi_{B}(z_{t}),$$
$$z_{t+1}= z_{t} \hspace{2mm} if \hspace{2mm} z_{t}=0  \hspace{2mm} or \hspace{2mm} z_{t}=n.$$

Now mutations are added. Players may switch to new strategies with the probability $\epsilon$. 
It is easy to see that for any two states of the population, there is a positive probability
of the transition between them in some finite number of time steps. 
We have therefore obtained an ergodic Markov chain with $n+1$ states 
and a unique stationary measure which we denote by $\mu^{\epsilon}_{n}.$
Kandori, Mailath, and Rob proved that the risk-dominant strategy $B$ is stochastically stable \cite{kmr}
\begin{theo}
$\lim_{\epsilon \rightarrow 0}\mu^{\epsilon}_{n}(0)=1$
\end{theo}
This means that in the long run, in the limit of no mutations, all players play $B$. 

The general set up in the Robson-Vega-Redondo model \cite{rvr} is the same.
However, individuals are paired only once at every time step and play only one game
before a selection process takes place. Let $p_{t}$ denote the random variable
which describes the number of cross-pairings, i.e. the number of pairs 
of matched individuals playing different strategies at the time $t$.
Let us notice that $p_{t}$ depends on $z_{t}$. For a given realization of  $p_{t}$ and $z_{t}$,
mean payoffs obtained by each strategy are as follows:
\begin{equation}
\tilde{\pi}_{A}(z_{t},p_{t})=\frac{a(z_{t}-p_{t})+bp_{t}}{z_{t}},
\end{equation} 
$$\tilde{\pi}_{B}(z_{t},p_{t})=\frac{cp_{t}+d(n-z_{t}-p_{t})}{n-z_{t}},$$
provided $0<z_{t}<n$. Robson and Vega-Redondo showed that the payoff-dominant strategy 
is stochastically stable \cite{rvr}. 
\begin{theo}
$\lim_{\epsilon \rightarrow 0}\mu^{\epsilon}_{n}(n)=1$
\end{theo}

We will outline their proof.

First of all, one can show that there exists $k$ such that 
if $n$ is large enough and $z_{t} \geq k$, 
then there is a positive probability (a certain realization of $p_{t}$)
that after a finite number of steps of the mutation-free selection dynamics,
all players will play $A$. Likewise, if $z_{t} <k$ (for any $k \geq 1$), 
then if the number of players is large enough, then after a finite number 
of steps of the mutation-free selection dynamics 
all players will play $B$. In other words, $z=0$ and $z=n$ 
are the only absorbing states of the mutation-free dynamics.
Moreover, if $n$ is large enough, then if $z_{t} \geq n-k$, then the mean payoff obtained
by $A$ is always (for any realization of $p_{t}$) bigger than the mean payoff obtained by $B$
(in the worst case all $B$-players play with $A$-players). Therefore the size of the basin 
of attraction of the state $z=0$ is at most $n-k-1$ and that of $z=n$ is at least $n-k$. 
Observe that mutation-free dynamics is not deterministic ($p_{t}$ describes the random matching)
and therefore basins of attraction may overlap. It follows that the system needs at least $k+1$ 
mutations to evolve from $z=n$ to $z=0$ and at most $k$ mutations to evolve from $z=0$ to $z=n$. 
Now using the tree representation of stationary states, Robson and Vega-Redondo finish the proof 
and show that the efficient strategy is stochastically stable. 
\vspace{2mm}

However, as outlined above, their proof requires the number of players to be sufficiently large. 
We will now show that a risk-dominant strategy is stochastically stable 
if the number of players is below certain value which can be arbitrarily large.

\begin{theo}
If $n<\frac{2a-c-b}{a-c}$, then the risk-dominant strategy $B$ is stochastically stable
in the case of random matching of players.
\end{theo}      

\noindent {\bf Proof:} If the population consists of only one $B$-player and $n-1$ $A$-players 
and if $c>[a(n-2)+b]/(n-1)$, that is $n< (2a-c-b)/(a-c)$, then  $\tilde{\pi}_{B}> \tilde{\pi}_{A}.$ 
It means that one needs only one mutation to evolve from $z=n$ to $z=0.$ 
It is easy to see that two mutations are necessary to evolve from $z=0$ to $z=n.$ 
\vspace{2mm}

To see stochastically stable states, we need to take the limit of no mutations. 
We will now examine the long-run behavior of the Robson-Vega-Redondo model 
for a fixed level of mutations in the limit of the infinite number of players.

Now we will analyze the extreme case of the selection rule (53) - 
a sequential dynamics where in one time unit only one player can change his strategy. 
Although our dynamics is discrete in time, it captures the essential features
of continuous-time models in which every player has an exponentially
distributed waiting time to a moment of a revision opportunity. 
Probability that two or more players revise their strategies 
at the same time is therefore equal to zero - this is an example 
of a birth and death process. 

The number of $A$-players in the population may increase by one
in $t+1$, if a $B$-player is chosen in $t$ which happens with 
the probability $(n-z_{t})/n$. Analogously, the number of $B$-players 
in the population may increase by one in $t+1$, if an $A$-player 
is chosen in $t$ which happens with the probability $(z_{t})/n$. 

The player who has a revision opportunity chooses in $t+1$ 
with the probability $1-\epsilon$ the strategy with a higher average payoff in $t$ 
and the other one with the probability $\epsilon$.  Let 
$$r(k)=P(\tilde{\pi}_{A}(z_{t},p_{t}) > \tilde{\pi}_{B}(z_{t},p_{t}))
\; \; and \; \; l(k)=P(\tilde{\pi}_{A}(z_{t},p_{t}) < \tilde{\pi}_{B}(z_{t},p_{t})).$$
The sequential dynamics is described by the following transition probabilities:
\vspace{1mm}

if $z_{t}=0 $, then $z_{t+1}=1 $ with the probability $\epsilon$
and $z_{t+1}=0 $ with the probability $1-\epsilon$,
\vspace{1mm}

if $z_{t}=n $, then $z_{t+1}=n-1 $ with the probability $\epsilon$
and $z_{t+1}=n $ with the probability $1-\epsilon$,
\vspace{1mm}

if $z_{t} \neq 0,n$, then $z_{t+1}= z_{t} +1$ with the probability

$$r(k)\frac{n-z_{t}}{n}(1-\epsilon)+(1-r(k))\frac{n-z_{t}}{n}\epsilon$$

and $z_{t+1}= z_{t} -1$ with the probability

$$l(k)\frac{z_{t}}{n}(1-\epsilon)+(1-l(k))\frac{z_{t}}{n}\epsilon.$$

In the dynamics intermediate between the parallel (where all individuals can revise their strategies 
at the same time) and the sequential one, each individual has a revision opportunity 
with the same probability $\tau <1$ during the time interval of the lenght $1$. 
For a fixed $\epsilon$ and an arbitrarily large but fixed $n$, we consider the limit of the continuous time, 
$\tau \rightarrow 0$, and show that the limiting behaviour is already obtained for a sufficiently small $\tau$, 
namely  $\tau  < \epsilon /n^{3}$. 
 
For an interesting discussion on the importance of the order of taking different limits 
$(\tau  \rightarrow 0, n \rightarrow \infty,$ and $\epsilon \rightarrow 0)$
in evolutionary models (especially in the Aspiration and Imitation model) 
see Samuelson \cite{samuel}.

In the intermediate dynamics, instead of $(n-z_{t})/n$ and 
$z_{t}/n$ probabilities we have more involved combinatorial 
factors. In order to get rid of these inconvenient factors, we will enlarge the state space
of the population. The state space $\Omega^{'}$ is the set of all configurations 
of players, that is all possible assignments of strategies to individual players.
Therefore, a state $z_{t}=k$ in $\Omega$ consists of 
$\left( \begin{array}{c} n \\ k \end{array} \right)$ 
states in $\Omega^{'}$. Observe that the sequential dynamics on $\Omega^{'}$ 
is not anymore a birth and death process. 
However, we are able to treat both dynamics in the same framework.

We showed in \cite{population} that for any arbitrarily low but fixed level of mutation, 
if the number of players is large enough, then in the long run only 
a small fraction of the population plays the payoff-dominant strategy. 
Smaller the mutation level is, fewer players use the payoff-dominant strategy.  
\vspace{1mm}

The following two theorems were proven in \cite{population}. 

\begin{theo}
In the sequential dynamics, for any $\delta >0$ 
and $\beta >0$ there exist $\epsilon(\delta, \beta)$
and $n(\epsilon)$ such that for any $n > n(\epsilon)$
$$\mu_{n}^{\epsilon}(z \leq \beta n) > 1- \delta.$$
\end{theo}

\begin{theo}
In the intermediate dynamics dynamics, for any $\delta >0$ 
and $\beta >0$ there exist $\epsilon(\delta, \beta)$
and $n(\epsilon)$ such that for any $n > n(\epsilon)$ and $\tau< \frac{\epsilon}{n^{3}}$
$$\mu_{n}^{\epsilon}(z \leq \beta n) > 1- \delta.$$
\end{theo}

We can combine Theorems 9, 10, and 12 and obtain \cite{banach}

\begin{theo}
In the intermediate dynamics, for any $\delta >0$ and $\beta >0$ there exists $\epsilon(\delta, \beta)$
such that, for all $\epsilon < \epsilon(\delta, \beta)$, there exist 
$n_{1} < n_{2} < n_{3}(\epsilon) < n_{4}(\epsilon)$ such that 

if $n < n_{1}=\frac{2a-c-b}{a-c}$, then $\mu_{n}^{\epsilon}(z =0) > 1- \delta,$

if $n_{2} < n < n_{3}(\epsilon)$, then $\mu_{n}^{\epsilon}(z = n) > 1- \delta,$

if $n > n_{4}(\epsilon)$ and $\tau< \epsilon/n^{3}$, 
then $\mu_{n}^{\epsilon}(z \leq \beta n) > 1- \delta$.
\end{theo}

\noindent Small $\tau$ means that our dynamics is close to the sequential one.
We have that $n_{3}(\epsilon), n_{4}(\epsilon), n_{3}(\epsilon)-n_{2}$, 
and $n_{4}(\epsilon)-n_{3}(\epsilon) \rightarrow \infty$ when $\epsilon \rightarrow 0$. 
\vspace{2mm}

It follows from Theorem 13 that the population of players undergoes several 
{\bf equilibrium transitions}. First of all, for a fixed noise level, when the number of players increases, 
the population switches from a $B$-equilibrium, where most of the individuals 
play the strategy $B$, to an $A$-equilibrium and then back to $B$ one. 
We know that if $n>n_{2}$, then $z=n$ is stochastically stable.
Therefore, for any fixed number of players, $n>n_{4}(\epsilon)$, 
when the noise level decreases, the population undergoes a transition from 
a $B$-equilibrium to $A$ one. We see that in order to study the long-run behaviour 
of stochastic population dynamics, we should estimate the relevant parameters 
to be sure what limiting procedures are appropriate in specific examples.  

Let us note that the above theorems concern an ensemble of states,
not an individual one. In the limit of the infinite number of players, 
that is the infinite number of states, every single state has zero probability
in the stationary state. It is an ensemble of states that might be stable \cite{statmech,physica}.
The concept of ensemble stability will be discussed in Chapter 9.

\section{Spatial games with local interactions}
\subsection{Nash configurations and stochastic dynamics}
In spatial games, players are located on vertices of certain graphs 
and they interact only with their neighbours; see for example
\cite{nowak0,nowak1,nowak2,blume1,ellis1,young2,ellis2,linnor,doebeli1,doebeli2,szabo1,szabo2, szabo6,hauert1,doebeli3,hauert2,hauert3}
and a recent review paper \cite{szabo8} and references therein. 

Let $\Lambda$ be a finite subset of the simple lattice ${\bf Z}^{d}$.
Every site of $\Lambda$ is occupied by one player who
has at his disposal one of $m$ different pure strategies. 
Let $S$ be the set of strategies, then $\Omega_{\Lambda}=S^{\Lambda}$ is the space
of all possible configurations of players, that is all possible assignments 
of strategies to individual players. For every $i \in \Lambda$, 
$X_{i}$ is a strategy of the $i-$th player in the configuration 
$X \in \Omega_{\Lambda}$ and $X_{-i}$ denotes strategies of all remaining players; 
$X$ therefore can be represented as the pair $(X_{i},X_{-i})$. 
Every player interacts only with his nearest neighbours and his payoff 
is the sum of the payoffs resulting from individual plays.
We assume that he has to use the same strategy for all neighbours. 
Let $N_{i}$ denote the neighbourhood of the $i-$th player. 
For the nearest-neighbour interaction we have $N_{i}=\{j; |j-i|=1\}$,
where $|i-j|$ is the distance between $i$ and $j$.
For $X \in \Omega_{\Lambda}$ we denote by $\nu_{i}(X)$ the payoff 
of the $i-$th player in the configuration $X$:
\begin{equation}
\nu_{i}(X)=\sum_{j \in N_{i}}U(X_{i}, X_{j}),
\end{equation}
where $U$ is a $m \times m$ matrix of payoffs of a two-player symmetric game with $m$ pure strategies.

\begin{defi}
$X \in \Omega_{\Lambda}$ is a {\bf Nash configuration} if for every $i \in \Lambda$
and $Y_{i} \in S$, 

$$\nu_{i}(X_{i},X_{-i}) \geq \nu_{i}(Y_{i},X_{-i})$$
\end{defi}

Here we will discuss only coordination games,
where there are $m$ pure symmetric Nash equilibria and therefore $m$ homogeneous
Nash configurations, where all players play the same strategy. 

In the Stag-hunt game in Example 1, we have two homogeneous Nash configurations, $X^{St}$ and $X^{H}$, 
where all individuals play $St$ or $H$ respectively.

We describe now the sequential deterministic dynamics of the {\bf best-response rule}. 
Namely, at each discrete moment of time $t=1,2,...$, a randomly chosen player may update 
his strategy. He simply adopts the strategy, $X_{i}^{t+1}$, which gives him 
the maximal total payoff $\nu_{i}(X_{i}^{t+1}, X^{t}_{-i})$ 
for given $X^{t}_{-i}$, a configuration of strategies 
of remaining players at the time $t$. 

Now we allow players to make mistakes, that is they may not choose best responses. 
We will discuss two types of such stochastic dynamics. In the first one, the so-called 
{\bf perturbed best response}, a player follows the best-response rule with probability $1-\epsilon$ 
(in case of more than one best-response strategy he chooses 
randomly one of them) and with probability $\epsilon$
he makes a mistake and chooses randomly one of the remaining strategies.
The probability of mistakes (or the noise level) is state-independent here.

In the so called {\bf log-linear dynamics}, the probability of chosing by the $i-$th player
the strategy $X_{i}^{t+1}$ at the time $t+1$ decreases with the loss of the payoff 
and is given by the following conditional probability:

\begin{equation}
p_{i}^{\epsilon}(X_{i}^{t+1}|X_{-i}^{t})=
\frac{e^{\frac{1}{\epsilon}\nu_{i}( X_{i}^{t+1},X_{-i}^{t})}}{\sum_{Y_{i} \in S}
e^{\frac{1}{\epsilon}\nu_{i}(Y_{i},X_{-i}^{t})}},
\end{equation}

Let us observe that if $\epsilon \rightarrow 0$, 
$p_{i}^{\epsilon}$ converges pointwise to the best-response rule.
Both stochastic dynamics are examples of ergodic Markov chains 
with $|S^{\Lambda}|$ states. Therefore they have unique stationary states 
denoted by $\mu_{\Lambda}^{\epsilon}$. 

Stationary states of the log-linear dynamics can be explicitly constructed 
for the so-called potential games. A game is called 
a {\bf potential game} if its payoff matrix can be changed 
to a symmetric one by adding payoffs to its columns \cite{mon}. 
As we know, such a payoff transformation does not change strategic 
character of the game, in particular it does not change the set of its
Nash equilibria. More formally, we have the following definition.

\begin{defi}
A two-player symmetric game with a payoff matrix $U$ \\ 
\noindent is a {\bf potential game} if there exists a symmetric matrix $V$, \\
\noindent called a potential of the game, 
such that for any three strategies $A, B, C \in S$
\begin{equation}
U(A,C)-U(B,C)=V(A,C)-V(B,C).
\end{equation} 
\end{defi}

It is easy to see that every game with two strategies 
has a potential $V$ with $V(A,A)=a-c$, $V(B,B)=d-b$,
and $V(A,B)=V(B,A)=0.$ It follows that an equilibrium is risk-dominant 
if and only if it has a bigger potential.
\vspace{1mm}

For players on a lattice, for any $X \in \Omega_{\Lambda}$,
\begin{equation}
V(X)=\sum_{(i,j) \subset \Lambda} V(X_{i},X_{j})
\end{equation}

is then the potential of the configuration $X$.
\vspace{2mm}

For the sequential log-linear dynamics of potential games, 
one can explicitely construct stationary measures \cite{young2}. 

We begin by the following general definition concerning a Markov chain 
with a state space $\Omega$ and a transition matrix $P$. 

\begin{defi} 
A measure $\mu$ on $\Omega$ satisfies a {\bf detailed balance condition} if
$$\mu(X)P_{XY}=\mu(Y)P_{YX}$$
for every $X,Y \in \Omega$ 
\end{defi} 

\noindent {\bf Lemma} 
\vspace{2mm}

{\em If $\mu$ satisfies the detailed balance condition then it is a stationary measure}
\vspace{2mm}

\noindent {\bf Proof:} 

$$\sum_{X \in \Omega}\mu(X)P_{XY}=\sum_{X \in \Omega} \mu(Y)P_{YX}= \mu(Y)$$

The following theorem is due Peyton Young \cite{young2}. We will present here his proof. 

\begin{theo}
The stationary measure of the sequential log-linear dynamics 
in a game with the potential $V$ is given by

\begin{equation}
\mu^{\epsilon}_{\Lambda}(X)=\frac{e^{\frac{1}{\epsilon}V(X)}}
{\sum_{Z \in \Omega_{\Lambda}}e^{\frac{1}{\epsilon}V(Z)}}.
\end{equation}
\end{theo}
\noindent {\bf Proof:}

We will show that $\mu^{\epsilon}_{\Lambda}$ in (59) satisfies the detailed balance condition. 
Let us notice that in the sequential dynamics, $P_{XY}=0$ unless $X=Y$ or $Y$ differs fom $X$
at one lattice site only, say $i \in \Lambda$. 

\noindent Let
 
$$\lambda =\frac{1}{|\Lambda|}\frac{1}{\sum_{Z \in \Omega_{\Lambda}}e^{\frac{1}{\epsilon}V(Z)}}\frac{1}{\sum_{Z_{i} 
\in S}e^{\frac{1}{\epsilon}\sum_{j \in N_{i}}U(Z_{i},X_{j})}}$$

\noindent Then

$$\mu^{\epsilon}_{\Lambda}(X)P_{XY}=\lambda e^{\frac{1}{\epsilon}(\sum_{(h,k) \subset \Lambda}
V(X_{h},X_{k})+\sum_{j \in N_{i}}U(Y_{i},X_{j}))}$$

$$=\lambda e^{\frac{1}{\epsilon}(\sum_{(h,k) \subset \Lambda}
V(X_{h},X_{k})+\sum_{j \in N_{i}}(U(X_{i},X_{j})-V(X_{i},X_{j})+V(Y_{i},X_{j})))}$$

$$=\lambda e^{\frac{1}{\epsilon}(\sum_{(h,k) \subset \Lambda}
V(Y_{h},Y_{k})+\sum_{j \in N_{i}}U(X_{i},X_{j}))}=\mu^{\epsilon}_{\Lambda}(Y)P_{YX}.$$
\vspace{3mm}

We may now explicitly perform the limit $\epsilon \rightarrow 0$ in (59).
In the Stag-hunt game, $X^{H}$ has a bigger potential than $X^{St}$ 
so $\lim_{\epsilon \rightarrow 0}\mu_{\Lambda}^{\epsilon}(X^{H})=1 $  
hence $X^{H}$ is stochastically stable (we also say that $H$ is stochastically stable).
\vspace{2mm}

The concept of a Nash configuration in spatial games is very similar to the concept 
of a ground-state configuration in lattice-gas models of interacting particles.
We will discuss similarities and differences between these two systems of interacting entities 
in the next section.

\subsection{Ground states and Nash configurations}

We will present here one of the basic models of interacting particles.
In classical lattice-gas models, particles occupy lattice sites and interact
only with their neighbours. The fundamental concept
is that of a ground-state configuration. It can be formulated conveniently
in the limit of an infinite lattice (the infinite number of particles).
Let us assume that every site of the ${\bf Z}^{d}$ lattice can be occupied 
by one of $m$ different particles. An infinite-lattice configuration 
is an assignment of particles to lattice sites, i.e. an element of $\Omega =
\{1,...,m\}^{{\bf Z}^{d}}$. If $X \in \Omega$ and $i \in  {\bf
Z}^{d}$, then we denote by $X_{i}$ a restriction of $X$ to $i$.
We will assume here that only nearest-neighbour particles interact.
The energy of their interaction is given by a symmetric $m \times m$ matrix $V$.
An element $V(A,B)$ is the interaction energy of two nearest-neighbour
particles of the type $A$ and $B$. The total energy of a system 
in the configuration $X$ in a finite region $\Lambda \subset {\bf Z}^{d}$ 
can be then written as 
\begin{equation}
H_{\Lambda}(X)=\sum_{(i,j) \subset \Lambda} V(X_{i},X_{j}).
\end{equation}

$Y$ is a {\bf local excitation} of $X$, $Y \sim X$, $Y,X \in
\Omega$ , if there exists a finite $\Lambda \subset {\bf Z}^{d}$
such that $X = Y$ outside $\Lambda.$ 

For $Y \sim X$, the {\bf relative energy} is defined by 
\begin{equation}
H(Y,X)=\sum_{(i,j) \in {\bf Z}^{d}} (V(Y_{i},Y_{j})-V(X_{i},X_{j})),
\end{equation}
where the summation is with respect to pairs of nearest neighbours
on ${\bf Z}^{d}$. Observe that this is the finite sum; the energy difference
between $Y$ and $X$ is equal to zero outside some finite $\Lambda$.

\begin{defi}
$X \in \Omega$ is a {\bf ground-state configuration} of $V$ if 
$$H(Y,X) \geq 0 \; \; for \; \; any \; \; Y \sim X.$$  
\end{defi}
That is, we cannot lower the energy of a ground-state
configuration by changing it locally.

The energy density $e(X)$ of a configuration $X$ is 
\begin{equation}
e(X)=\liminf_{\Lambda \rightarrow {\bf Z}^{2}}
\frac{H_{\Lambda}(X)}{|\Lambda|},
\end{equation}

where $|\Lambda|$ is the number of lattice sites in $\Lambda$. 
\vspace{1mm}

It can be shown that any  ground-state configuration has the minimal
energy density \cite{sinai}. It means that local conditions present in the definition 
of a ground-state configuration force the global minimization of the energy density.

We see that the concept of a ground-state configuration is very similar
to that of a Nash configuration. We have to identify
particles with agents, types of particles with strategies
and instead of minimizing interaction energies we should maximize payoffs.
There are however profound differences. First of all, 
ground-state configurations can be defined only for symmetric matrices; 
an interaction energy is assigned to a pair of particles, payoffs are assigned 
to individual players and may be different for each of them.
Ground-state configurations are stable with respect to all local changes, 
Nash configurations are stable only with respect to one-player changes.
It means that for the same symmetric matrix $U$, there may exist a configuration 
which is a Nash configuration but not a ground-state configuration 
for the interaction matrix $-U$. The simplest example is given by the following matrix:
\eject

\noindent {\bf Example 5}

\hspace{23mm} A \hspace{4mm} B   
\vspace{1mm}

\hspace{15mm} A \hspace{3mm} 2 \hspace{5mm} 0 

U = \hspace{6mm} 

\hspace{15mm} B \hspace{3mm} 0 \hspace{5mm} 1 
\vspace{3mm}

\noindent $(A,A)$ and $(B,B)$ are Nash configurations for a system consisting 
of two players but only $(A,A)$ is a ground-state configuration for $V=-U.$  
We may therefore consider the concept of a ground-state configuration 
as a refinement of a Nash equilibrium.

For any classical lattice-gas model there exists at least one 
ground-state configuration. This can be seen in the following way.
We start with an arbitrary configuration. If it cannot be changed locally
to decrease its energy it is already a ground-state configuration.
Otherwise we may change it locally and decrease the energy of the system.
If our system is finite, then after a finite number of steps we arrive at a
ground-state configuration; at every step we decrease the energy of the system
and for every finite system its possible energies form a finite set.
For an infinite system, we have to proceed ad infinitum converging
to a ground-state configuration (this follows from the compactness of $\Omega$ 
in the product of discrete topologies). Game models are different. 
It may happen that a game with a nonsymmetric payoff matrix may not posess 
a Nash configuration. The classical example is that of the Rock-Scissors-Paper game.
One may show that this game dos not have any Nash configurations on ${\bf Z}$
and ${\bf Z}^{2}$ but many Nash configurations on the triangular lattice.

In short, ground-state configurations minimize the total energy of a particle system, 
Nash configurations do not necessarily maximize the total payoff of a population.

Ground-state configuration is an equilibrium concept for
systems of interacting particles at zero temperature. For positive temperatures,
we must take into account fluctuations caused by thermal motions of particles.
Equilibrium behaviour of the system results then from the competition between
its energy $V$ and entropy $S$ (which measures the number of configurations corresponding 
to a macroscopic state), i.e. the minimization of its free energy $F=V-TS$, 
where $T$ is the temperature of the system - a measure of thermal motions. 
At the zero temperature, $T=0$, the minimization of the free energy reduces to the minimization of the energy.
This zero-temperature limit looks very similar to the zero-noise limit present 
in the definition of the stochastic stability. Equilibrium behaviour of a system 
of interacting particles can be described by specifying probabilities of occurence 
for all particle configurations. More formally, it is described 
by a Gibbs state (see \cite{geo} and references therein).

We construct it in the following way. Let
$\Lambda$ be a finite subset of ${\bf Z}^{d}$ and  $\rho^{T}_{\Lambda}$ 
the following probability mass function on
$\Omega_{\Lambda}=(1,...,m)^{\Lambda}$: 
\begin{equation}
\rho_{\Lambda}^{T}(X)=(1/Z^{T}_{\Lambda})\exp(-H_{\Lambda}(X)/T),
\end{equation}
for every $X \in \Omega_{\Lambda}$, where
\begin{equation}
Z^{T}_{\Lambda}=\sum_{X \in \Omega_{\Lambda}}\exp(-H_{\Lambda}(X)/T)
\end{equation}
is a normalizing factor.
\vspace{1mm}

We define a {\bf Gibbs state} $\rho^{T}$ as a limit of $\rho^{T}_{\Lambda}$ as
$\Lambda \rightarrow {\bf Z}^{d}$. One can prove
that a limit of a translation-invariant Gibbs state for a given interaction
as $T \rightarrow 0$ is a measure supported by ground-state configurations.
One of the fundamental problems of statistical mechanics is a characterization
of low-temperature Gibbs states for given interactions between particles.

Let us observe that the finite-volume Gibbs state in (63) is equal 
to stationary state $\mu^{\epsilon}_{\Lambda}$ in (59) if we identify $T$ with $\epsilon$
and $V \rightarrow -V$.

\subsection{Ensemble stability}

The concept of stochastic stability involves individual configurations of players. 
In the zero-noise limit, a stationary state is usually concentrated on one 
or at most few configurations. However, for a low but fixed noise and
for a sufficiently large number of players, the probability of any individual configuration 
of players is practically zero. The stationary measure, however, may be highly 
concentrated on an ensemble consisting of one Nash configuration and its small 
perturbations, i.e. configurations where most players use the same strategy. 
Such configurations have relatively high probability in the stationary measure. 
We call such configurations ensemble stable. Let $\mu^{\epsilon}_{\Lambda}$ 
be a stationary measure.

\begin{defi}
$X \in \Omega_{\Lambda}$ is {\bf $\gamma$-ensemble stable}
if $\mu_{\Lambda}^{\epsilon}(Y \in \Omega_{\Lambda};Y_{i} \neq X_{i}) < \gamma$
for any $i \in \Lambda$ if $\Lambda \supset \Lambda(\gamma)$ for some $\Lambda(\gamma)$. 
\end{defi}

\begin{defi}
$X \in \Omega_{\Lambda}$ is {\bf low-noise ensemble stable}
if for every $\gamma>0$ there exists $\epsilon(\gamma)$ such that if
$\epsilon<\epsilon(\gamma)$, then $X$ is $\gamma$-ensemble stable.
\end{defi}

If $X$ is $\gamma$-ensemble stable with $\gamma$ close to zero, then the ensemble 
consisting of $X$ and configurations which are different from $X$ at at most few sites has 
the probability close to one in the stationary measure. It does not follow, however, 
that $X$ is necessarily low-noise ensemble or stochastically stable as it happens 
in examples presented below \cite{statmech}.
\eject

\noindent {\bf Example 6}
\vspace{2mm}

\noindent Players are located on a finite subset $\Lambda$ of ${\bf Z}^{2}$ 
(with periodic boundary conditions) and interact with their 
four nearest neighbours. They have at their disposal three pure strategies: 
$A, B,$ and $C$. The payoffs are given by the following symmetric matrix:
\vspace{5mm}

\hspace{23mm} A  \hspace{2mm} B \hspace{2mm} C  

\hspace{15mm} A  \hspace{2mm} 1.5  \hspace{2mm} 0 \hspace{2mm} 1

U = \hspace{7mm} B \hspace{3mm} 0  \hspace{4mm} 2 \hspace{2mm} 1

\hspace{15mm} C \hspace{3mm} 1  \hspace{4mm} 1 \hspace{2mm} 2
\vspace{2mm}

Our game has three Nash equilibria: $(A,A), (B,B)$, and $(C,C)$,
and the corresponding spatial game has three homogeneous Nash configurations:
$X^{A}, X^{B}$, and $X^{C}$, where all individuals are assigned the same strategy. 
Let us notice that $X^{B}$ and $X^{C}$ have the maximal
payoff in every finite volume and therefore they are ground-state configurations for $-U$
and $X^{A}$ is not.

The unique stationary measure of the log-linear dynamics (56) is  is given by (59) with $U=V$
which is a finite-volume Gibbs state (63) with $V$ replaced by $-U$ and $T$ by $\epsilon$.
We have  

$$\sum_{(i,j)\subset \Lambda}U(X^{k}_{i},X^{k}_{j})-\sum_{(i,j)\in \Lambda}U(Y_{i},Y_{j})>0,$$
for every $Y \neq X^{B} \; and \; X^{C}$, $k=B,C$, and 
$$\sum_{(i,j)\subset \Lambda}U(X^{B}_{i},X^{B}_{j})=\sum_{(i,j)\subset \Lambda}U(X^{C}_{i},X^{C}_{j}).$$

It follows that $\lim_{\epsilon \rightarrow 0}\mu_{\Lambda}^{\epsilon}(X^{k})=1/2$, 
for $k=B, C $ so $X^{B}$ and $X^{C}$ are stochastically stable.
Let us investigate the long-run behaviour of our system for large $\Lambda$, 
that is for a large number of players. 

Observe that 
$$\lim_{\Lambda \rightarrow {\bf Z}^{2}}\mu_{\Lambda}^{\epsilon}(X)=0$$ 

for every $X \in \Omega = S^{{\bf Z}^{2}}$.

Therefore, for a large $\Lambda$ we may only observe,
with reasonably positive frequencies, ensembles of configurations 
and not particular configurations. We will be interested in ensembles 
which consist of a Nash configuration and its small perturbations, 
that is configurations, where most players use the same strategy. 
We perform first the limit $\Lambda \rightarrow {\bf Z}^{2}$
and obtain an infinite-volume Gibbs state in the temperature $T=\epsilon$,
\begin{equation}
\mu^{\epsilon} = \lim_{\Lambda \rightarrow {\bf Z}^{2}}\mu_{\Lambda}^{\epsilon}.
\end{equation}

In order to investigate the stationary state of our example, we will apply a technique developed 
by Bricmont and Slawny \cite{brsl1,brsl2}. They studied low-temperature stability of the so-called dominant 
ground-state configurations. It follows from their results that

\begin{equation}
\mu^{\epsilon}(X_{i}=C)>1-\delta(\epsilon)  
\end{equation}
for any $i \in {\bf Z}^{2}$ and $\delta(\epsilon) \rightarrow 0$ as $\epsilon \rightarrow 0$ \cite{statmech}. 
\vspace{2mm}

The following theorem is a simple consequence of (66).
\begin{theo}
$X^{C}$ is low-noise ensemble stable.
\end{theo} 

We see that for any low but fixed $\epsilon$, if the number of players is large enough,
then in the long run, almost all players use $C$ strategy. 
On the other hand, if for any fixed number of players, $\epsilon$ is lowered substantially,
then B and C appear with frequencies close to $1/2$.

Let us sketch briefly the reason of such a behavior.
While it is true that both $X^{B}$ and $X^{C}$ have the same potential 
which is the half of the payoff of the whole system (it plays the role 
of the total energy of a system of interacting particles), 
the $X^{C}$ Nash configuration has more lowest-cost excitations. 
Namely, one player can change its strategy and switch to either
$A$ or $B$ and the potential will decrease by $4$ units. Players 
in the $X^{B}$ Nash configuration have only one possibility, 
that is to switch to $C$; switching to $A$ decreases the potential by $8$. 
Now, the probability of the occurrence of any configuration in the Gibbs state 
(which is the stationary state of our stochastic dynamics) 
depends on the potential in an exponential way. 
One can prove that the probability of the ensemble consisting of the $X^{C}$ Nash 
configuration and configurations which are different from it 
at few sites only is much bigger than the probability of the analogous
$X^{B}$-ensemble. It follows from the fact that the $X^{C}$-ensemble 
has many more configurations than the $X^{B}$-ensemble. On the other hand,
configurations which are outside $X^{B}$ and $X^{C}$-ensembles 
appear with exponentially small probabilities. It means that for large enough systems 
(and small but not extremely small $\epsilon$) we observe in the stationary state the $X^{C}$ 
Nash configuration with perhaps few different strategies. The above argument was made 
into a rigorous proof for an infinite system of the closely related lattice-gas model 
(the Blume-Capel model) of interacting particles by Bricmont and Slawny in \cite{brsl1}.  

In the above example, $X^{B}$ and $X^{C}$ 
have the same total payoff but $X^{C}$ has more lowest-cost excitations
and therefore it is low-noise ensemble stable. We will now discuss the situation,
where $X^{C}$ has a smaller total payoff but nevertheless in the long run
$C$ is played with a frequency close to $1$ if the noise level is low but not extremely low. 
We will consider a family of games with the following payoff matrix:
\vspace{2mm}

\noindent {\bf Example 7}
\vspace{2mm}

\hspace{23mm} A  \hspace{5mm} B \hspace{5mm} C  

\hspace{15mm} A \hspace {3mm} 1.5  \hspace{3mm} 0 \hspace{6mm} 1

U = \hspace{7mm} B \hspace{4mm} 0  \hspace{3mm} $2+\alpha$ \hspace{1mm} 1

\hspace{15mm} C \hspace{3mm} 1  \hspace{6mm} 1 \hspace{5mm} 2,
\vspace{2mm}

where $\alpha>0$ so $B$ is both payoff and pairwise risk-dominant.
\vspace{2mm}

We are interested in the long-run behavior of our system for small positive $\alpha$
and low $\epsilon$. One may modify the proof of Theorem 15 and obtain 
the following theorem \cite{statmech}.
\begin{theo}
For every $\gamma>0$, there exist $\alpha(\gamma)$ and $\epsilon(\gamma)$ 
such that for every $0<\alpha<\alpha(\gamma)$, 
there exists $\epsilon(\alpha)$ such that for $\epsilon(\alpha)<\epsilon<\epsilon(\gamma)$, 
$X^{C}$ is $\gamma$-ensemble stable,
and for $0<\epsilon<\epsilon(\alpha)$, $X^{B}$ is $\gamma$-ensemble stable.
\end{theo}

Observe that for $\alpha=0$, both $X^{B}$ and $X^{C}$ are stochastically stable
(they appear with the frequency $1/2$ in the limit of zero noise) but $X^{C}$
is low-noise ensemble stable. For small $\alpha > 0$, $X^{B}$ is both stochastically
(it appears with the frequency $1$ in the limit of zero noise) and low-noise 
ensemble stable. However, for an intermediate noise $\epsilon(\alpha)<\epsilon<\epsilon(\gamma)$, 
if the number of players is large enough, then in the long run, 
almost all players use the strategy $C$ ($X^{C}$ is ensemble stable). 
If we lower $\epsilon$ below $\epsilon(\alpha)$, then almost all players start to use the strategy $B$. 
$\epsilon=\epsilon(\alpha)$ is the line of the first-order phase transition. In the thermodynamic limit, 
there exist two Gibbs states (equilibrium states) on this line. We may say that at $\epsilon=\epsilon(\alpha)$, 
the population of players undergoes a sharp {\bf equilibrium transition} from $C$ to $B$-behaviour. 

\subsection{Stochastic stability in non-potential games}

Let us now consider non-potential games with three strategies and three
symmetric Nash equilibria: $(A,A), (B,B)$, and $(C,C)$. Stationary measures 
of such games cannot be explicitly constructed. To find stochastically stable states
we will use here the tree representation of stationary measures described in Chapter 7. 
We will discuss some interesting examples. 
\vspace{2mm}

\noindent {\bf Example 8}
\vspace{2mm}

\noindent Players are located on a finite subset of the one-dimensional lattice 
${\bf Z}$ and interact with their nearest neighbours only. 
Denote by $n$ the number of players. For simplicity we will assume 
periodic boundary conditions, that is we will identify the $n+1$-th
player with the first one. In other words, the players are located on the circle.

The payoffs are given by the following  matrix:
\vspace{2mm}

\hspace{28mm} A \hspace{5mm} B \hspace {4mm} C

\hspace{15mm} A \hspace{5mm} $1+\alpha$  \hspace{3mm} 0 \hspace{3mm} 1.5

U = \hspace{7mm} B \hspace{9mm} 0  \hspace{6mm} 2 \hspace{4mm} 0

\hspace{15mm} C \hspace{9mm} 0  \hspace{6mm} 0 \hspace{4mm} 3
\vspace{5mm}

with $0< \alpha \leq 0.5$.
\vspace{2mm}

As before, we have three homogeneous Nash configurations: $X^{A}, X^{B}$,
and $X^{C}$. The log-linear and perturbed best-response dynamics for this game
were discussed in \cite{statphys}. 

Let us note that $X^{A}$, $X^{B}$, and $X^{C}$ are the only absorbing states 
of the noise-free dynamics. We begin with a stochastic dynamics with a state-independent noise. 
Let us consider first the case of $\alpha < 0.5$. 

\begin{theo}
If $0< \alpha< 0.5$, then $X^{C}$ is stochastically stable in the perturbed best-response dynamics. 
\end{theo} 

\noindent {\bf Proof:} It is easy to see that $q_{m}(X^{C})$ is of the order $\epsilon^{2}$,
$q_{m}(X^{B})$ is of the order $\epsilon^{\frac{n}{2}+1}$, and
$q_{m}(X^{A})$ is of the order $\epsilon^{n+2}$. 
\vspace{3mm}

Let us now consider the log-linear rule.

\begin{theo}
If $n<2+1/(0.5-\alpha)$, then $X^{B}$ is stochastically stable and
if $n>2+1/(0.5-\alpha)$, then $X^{C}$ is stochastically stable in the log-linear dynamics. 
\end{theo} 

{\bf Proof}: The following are maximal A-tree, B-tree, and C-tree:
$$B \rightarrow C \rightarrow A, \hspace{3mm} C \rightarrow A \rightarrow B,
\hspace{3mm} A \rightarrow B \rightarrow C,$$

where the probability of $A \rightarrow B$ is equal to
\begin{equation}
\frac{1}{1+1+e^{\frac{1}{\epsilon}(2+2\alpha)}}(\frac{1}{1+e^{-\frac{2}{\epsilon}}+e^{\frac{1}{\epsilon}(-1+\alpha)}})^{n-2} \frac{1}{1+e^{-\frac{4}{\epsilon}}+e^{-\frac{4}{\epsilon}}},
\end{equation}

the probability of $B \rightarrow C$ is equal to
\begin{equation}
\frac{1}{1+1+e^{\frac{4}{\epsilon}}}(\frac{1}{1+e^{-\frac{1}{\epsilon}}+e^{-\frac{1.5}{\epsilon}}})^{n-2}\frac{1}
{1+e^{-\frac{6}{\epsilon}}+e^{-\frac{3}{\epsilon}}},
\end{equation}

and the probability of $C \rightarrow A$ is equal to
\begin{equation}
\frac{1}{1+e^{-\frac{3}{\epsilon}}+e^{\frac{3}{\epsilon}}}(\frac{1}{1+e^{-\frac{1}{\epsilon}(2.5+\alpha)}
+e^{\frac{1}{\epsilon}(0.5-\alpha)}})^{n-2}\frac{1}{1+e^{-\frac{2}{\epsilon}(1+\alpha)}+e^{-\frac{2}{\epsilon}(1+\alpha)}},
\end{equation}

Let us observe that 
\begin{equation}
P_{B \rightarrow C \rightarrow A}= O(e^{-\frac{1}{\epsilon}(7+(0.5-\alpha)(n-2))}),
\end{equation}
\begin{equation}
P_{C \rightarrow A \rightarrow B}=
O(e^{-\frac{1}{\epsilon}(5+2\alpha+(0.5-\alpha)(n-2))}),
\end{equation}
\begin{equation}
P_{A \rightarrow B \rightarrow C}= O(e^{-\frac{1}{\epsilon}(6+2\alpha)}),
\end{equation}
where $\lim_{x \rightarrow 0}O(x)/x =1.$
\vspace{1mm}

Now if $n<2+1/(0.5-\alpha)$, then 
\begin{equation}
\lim_{\epsilon \rightarrow 0}\frac{q_{m}(X^{C})}{q_{m}(X^{B})}=
\lim_{\epsilon \rightarrow 0}\frac{P_{A \rightarrow B \rightarrow C}}
{P_{C \rightarrow A \rightarrow B}}=0
\end{equation}
which finishes the proof.
\vspace{2mm}

It follows that for a small enough $n$, $X^{B}$ is stochastically stable
and for a large enough $n$, $X^{C}$ is stochastically stable. 
We see that adding two players to the population 
may change the stochastic stability of Nash configurations. 
Let us also notice that the strategy $C$ is globally risk dominant. 
Nevertheless, it is not stochastically stable in the log-linear dynamics 
for a sufficiently small number of players. 
\vspace{1mm}

Let us now discuss the case of $\alpha=0.5$ \cite{statphys}.

\begin{theo}
If $\alpha=0.5$, then $X^{B}$ is stochastically stable for any $n$ in the log-linear dynamics. 
\end{theo}

{\bf Proof}: 
$$\lim_{\epsilon \rightarrow 0}\frac{q_{m}(X^{C})}{q_{m}(X^{B})}=
\lim_{\epsilon \rightarrow 0}\frac{e^{-\frac{4}{\epsilon}}e^{-\frac{3}{\epsilon}}}{(1/2)^{n-2}e^{-\frac{3}{\epsilon}}e^{-\frac{3}{\epsilon}}}=0.$$
\vspace{2mm}

\noindent $X^{B}$ is stochastically stable which means that for any fixed number of players, 
if the noise is sufficiently small, then in the long run we observe $B$ players 
with an arbitrarily high frequency. However, we conjecture that for any low but fixed noise, 
if the number of players is big enough, the stationary measure is concentrated on the $X^{C}$-ensemble.
We expect that $X^{C}$ is ensemble stable because its lowest-cost excitations 
occur with a probability of the order $e^{-\frac{3}{\epsilon}}$ and those from $X^{B}$ with a probability 
of the order $e^{-\frac{4}{\epsilon}}$.  We observe this phenomenon in Monte-Carlo simulations. 
\vspace{2mm}

\noindent {\bf Example 9}
\vspace{2mm}

\noindent Players are located on a finite subset $\Lambda$ of ${\bf Z}$ 
(with periodic boundary conditions) and interact with their 
two nearest neighbours. They have at their disposal three pure strategies: 
$A, B,$ and $C$. The payoffs are given by the following matrix \cite{statmech}:
\vspace{5mm}

\hspace{23mm} A  \hspace{2mm} B \hspace{2mm} C  

\hspace{15mm} A  \hspace{3mm} 3  \hspace{3mm} 0 \hspace{3mm} 2

U = \hspace{7mm} B \hspace{3mm} 2  \hspace{3mm} 2 \hspace{3mm} 0

\hspace{15mm} C \hspace{3mm} 0  \hspace{3mm} 0 \hspace{3mm} 3
\vspace{3mm}

\noindent Our game has three Nash equilibria: $(A,A), (B,B)$, and $(C,C)$.
Let us note that in pairwise comparisons, $B$ risk dominates $A$, 
$C$ dominates $B$ and $A$ dominates $C$. The corresponding spatial 
game has three homogeneous Nash configurations: $X^{A}, X^{B}$, and $X^{C}$. 
They are the only absorbing states of the noise-free best-response dynamics. 

\begin{theo}
$X^{B}$ is stochastically stable
\end{theo} 

{\bf Proof}: The following are maximal A-tree, B-tree, and C-tree:
$$B \rightarrow C \rightarrow A, \hspace{3mm} C \rightarrow A \rightarrow B,
\hspace{3mm} A \rightarrow B \rightarrow C.$$

Let us observe that
\begin{equation}
P_{B \rightarrow C \rightarrow A}= O(e^{-\frac{6}{\epsilon}}),
\end{equation}
\begin{equation}
P_{C \rightarrow A \rightarrow B}= O(e^{-\frac{4}{\epsilon}}),
\end{equation}
\begin{equation}
P_{A \rightarrow B \rightarrow C}= O(e^{-\frac{6}{\epsilon}}).
\end{equation}

\noindent The theorem follows from the tree characterization of stationary measures.
\vspace{3mm}

$X^{B}$ is stochastically stable because it is much more probable (for low $\epsilon$) 
to escape from $X^{A}$ and $X^{C}$ than from $X^{B}$. The relative payoffs 
of Nash configurations are not relevant here (in fact $X^{B}$ has the smallest payoff).   
Let us recall Example 7 of a potential game, where an ensemble-stable configuration
has more lowest-cost excitations. It is easier to escape 
from an ensemble-stable configuration than from other Nash configurations. 

Stochatic stability concerns single configurations in the zero-noise limit; 
ensemble stability concerns families of configurations 
in the limit of the infinite number of players. It is very important 
to investigate and compare these two concepts of stability in nonpotential games.  

Non-potential spatial games cannot be directly presented as systems of interacting
particles. They constitute a large family of interacting objects not thoroughly studied 
so far by methods statistical physics. Some partial results concerning stochastic stability 
of Nash equilibria in non-potential spatial games were obtained in 
\cite{ellis1,ellis2,blume1,physica,statphys}.

One may wish to say that $A$ risk dominates the other two strategies 
if it risk dominates them in pairwise comparisons. In Example 9, 
$B$ dominates $A$, $C$ dominates $B$, and finally $A$ dominates $C$. 
But even if we do not have such a cyclic relation of dominance, 
a strategy which is pairwise risk-dominant may not be stochastically stable 
as in the case of Example 8. 
A more relevant notion seems to be that of a global risk dominance \cite{mar}.
We say that $A$ is globally risk dominant if it is a best response
to a mixed strategy which assigns probability $1/2$ to $A$.
It was shown in \cite{ellis1,ellis2} that a global risk-dominant 
strategy is stochastically stable in some spatial games with local interactions. 

A different criterion for stochastic stability was developed by Blume
\cite{blume1}. He showed (using techniques of statistical mechanics) 
that in a game with $m$ strategies $A_{i}$ and $m$ symmetric Nash equilibria
$(A_{k},A_{k})$, $k=1,...,m$, $A_{1}$ is stochastically stable if
\begin{equation}
\min_{k>1}(U(A_{1},A_{1})-U(A_{k},A_{k})) > \max_{k>1}(U(A_{k},A_{k})-U(A_{1},A_{k})).
\end{equation}
We may observe that if $A_{1}$ satisfies the above condition, then it is pairwise
risk dominant. 

\subsection{Dominated strategies}

We say that a pure strategy is {\bf strictly dominated} by another (pure or mixed) strategy 
if it gives a player a lower payoff than the other one regardless of strategies 
chosen by his opponents.

\begin{defi}
$k \in S$ is strictly dominated by $y \in \Delta$ if $U_{i}(k,w_{-i}) < U_{i}(y,w_{-i})$ 
for every $w \in \Delta^{I}$. 
\end{defi}

Let us see that a strategy can be strictly dominated by a mixed strategy 
without being strictly dominated by any pure strategy in its support.
\vspace{3mm}

\noindent {\bf Example 10}
\vspace{2mm}

\hspace{23mm} A  \hspace{2mm} B \hspace{2mm} C  

\hspace{15mm} A  \hspace{3mm} 5  \hspace{3mm} 1 \hspace{3mm} 3

U = \hspace{7mm} B \hspace{3mm} 2  \hspace{3mm} 2 \hspace{3mm} 2

\hspace{15mm} C \hspace{3mm} 1  \hspace{3mm} 5 \hspace{3mm} 3
\vspace{3mm}

$B$ is strictly dominated by a mixed strategy assigning the probability $1/2$ both to $A$ and $C$ 
but is strictly dominated neither by $A$ nor by $C$.
\vspace{3mm}

It is easy to see that strictly dominated pure strategies cannot be present 
in the support of any Nash equilibrium. 

In the replicator dynamics (16), all strictly dominated pure strategies are wiped out in the long run
if all strategies are initially present \cite{akin,samzhang}.

\begin{theo}
If a pure strategy $k$ is strictly dominated, \\ then $\xi_{k}(t,x^{0}) \rightarrow_{t \rightarrow \infty} 0$
for any $x^{0} \in interior(\Delta)$. 
\end{theo}

Strictly dominated strategies should not be used by rational players and consequently 
we might think that their presence should not have any impact on the long-run behaviour of the population. 
We will show that in the best-reply dynamics, if we allow players to make mistakes, 
this may not be necessarily true. Let us consider the following game with a strictly 
dominated strategy and two symmetric Nash equilibria \cite{statmech}. 
\vspace{2mm}

\noindent {\bf Example 11}
\vspace{2mm}

\hspace{23mm} A  \hspace{5mm} B \hspace{5mm} C  

\hspace{15mm} A \hspace {3mm} 0  \hspace{5mm} 0.1 \hspace{4mm} 1

U = \hspace{7mm} B \hspace{2mm} 0.1  \hspace{1mm} $2+\alpha$ \hspace{1mm} 1.1

\hspace{15mm} C \hspace{2mm} 1.1 \hspace{4mm} 1.1 \hspace{3mm} 2,
\vspace{2mm}

where $\alpha>0$.
\vspace{3mm}

\noindent We see that strategy $A$ is strictly dominated by both $B$ and $C$, 
hence $X^{A}$ is not a Nash configuration. $X^{B}$ and $X^{C}$
are both Nash configurations but only $X^{B}$ is a ground-state 
configuration for $-U.$ In the absence of $A$,
$B$ is both payoff and risk-dominant and therefore is stochastically stable
and low-noise ensemble stable. Adding the strategy $A$ does not change dominance 
relations; $B$ is still payoff and pairwise risk dominant.
However, Example 11 fulfills all the assumptions of Theorem 16 and we get that 
$X^{C}$ is $\gamma$-ensemble stable at intermediate noise levels. 
The mere presence of a strictly dominated strategy $A$ changes the long-run behaviour 
of the population. 

Similar results were discussed by Myatt and Wallace \cite{wallace}. 
In their games, at every discrete moment of time, one of the players leaves 
the population and is replaced by another one who plays the best response. 
The new player calculates his best response with respect to his own payoff matrix 
which is the matrix of a common average payoff modified by a realization 
of some random variable with the zero mean. The noise does not appear in the game 
as a result of players' mistakes but is the effect of their idiosyncratic preferences. 
The authors then show that the presence of a strictly dominated strategy may change 
the stochastic stability of Nash equilibria. However, the reason for such a behavior
is different in their and in our models. In our model, it is relatively easy 
to get out of $X^{C}$ and this makes $X^{C}$ ensemble stable. Mayatt and Wallace introduce
a strictly dominated strategy in such a way that it is relatively easy to make a transition to it 
from a risk and payoff-dominant equilibrium and then with a high probability 
the population moves to a second Nash configuration which results in its stochastic stability.

This is exactly a mechanism present in Examples 8 and 9. 

\section{Review of other results}

We discussed the long-run behaviour of populations of interacting individuals
playing games. We have considered deterministic replicator dynamics and stochastic 
dynamics of finite populations. 
 
In spatial games, individuals are located on vertices of certain graphs 
and they interact only with their neighbours.  

In this paper, we considered only simple graphs - finite subsets of the regular $\bf Z$ 
or ${\bf Z}^{2}$ lattice. Recently there appeared many interesting results of evolutionary dynamics
on random graphs, Barabasi-Albert free-scale graphs, and small-world networks 
\cite{szabo3,szabo4,szabo5,szabo7,szabo8,santos1,santos2,santos3,antal}.
Especially the Prisoner's Dilemma was studied on such graphs and it was shown 
that their heterogeneity favors the cooperation in the population \cite{santos1,santos2,santos3,szabo8}.  

In well-mixed populations, individuals are randomly matched 
to play a game. The deterministic selection part of the dynamics ensures 
that if the mean payoff of a given strategy is bigger than the mean payoff of the other one, 
then the number of individuals playing the given strategy increases. 
In discrete moments of time, individuals produce offspring proportional to their payoffs. 
The total number of individuals is then scaled back to the previous value so the population size 
is constant. Individuals may mutate so the population may move against a selection pressure.
This is an example of a stochastic frequency-dependent Wright-Fisher process 
\cite{fisher1,fisher2,wright,burger,ewens}. 

There are also other stochastic dynamics of finite populations. The most important one 
is the Moran process \cite{moran,burger,ewens}. In this dynamics, at any time step a single individual
is chosen for reproduction with the probability proportional to his payoff, and then his offspring 
replaces the random chosen individual. It was showed recently that in the limits of the infinite population, 
the Moran process results in the replicator dynamics \cite{claussen1,claussen2}. 

The stochastic dynamics of finite populations has been extensively studied recently
\cite{nowak3,nowak4,doebeli3,nowak5,lieberman,ohtsuki1,ohtsuki2,ohtsuki3,traulsen1,traulsen2}.
The notion of an evolutionarily stable strategy for finite populations was introduced 
\cite{nowak3,nowak4,neill,wild,dostal,traulsen}.
One of the important quantity to calculate is the fixation probability of a given strategy. 
It is defined as the probability that a strategy introduced into a population by a single player 
will take over the whole population. Recently, Nowak et. al. \cite{nowak3} have formulated 
the following weak selection 1/3 law. In two-player games with two strategies, 
selection favors the strategy $A$ replacing $B$ if the fraction of A-players 
in the population for which the average payoff for the strategy $A$ is equal 
to the average payoff of the strategy $B$ if is smaller than $1/3$, 
i.e. the mixed Nash equilibrium for this game is smaller than $1/3$.  
The $1/3$ law was proven to hold both in the Moran \cite{nowak3,nowak4} 
and the Wright-Fisher process \cite{nowak5}.

In this review we discussed only two-player games. Multi-player games were studied recently 
in \cite{kim,broom,multi,physica,tplatk1,tplatk2,bulletin,tplatk3}.

We have not discussed at all population genetics in the context of game theory.
We refer to \cite{hof2,burger} for results and references.
\vspace{5mm}

\noindent {\bf Acknowledgments}: These lecture notes are based on the short course given 
in the Stefan Banach International Mathematical Center in the framework 
of the CIME Summer School ``From a Microscopic to a Macroscopic Description of Complex Systems" 
which was held in B\c{e}dlewo, Poland, 4-9 September 2006.

I would like to thank the Banach Center for a financial support to participate in this School
and the Ministry of Science and Higher Education for a financial support under the grant N201 023 31/2069.

I thank Martin Nowak and Arne Traulsen for useful comments.

\end{document}